\documentclass[aps,superscriptaddress,long,notitlepage,letterpaper,balancelastpage,nofootinbib,prd,floatfix,twocolumn]{revtex4}
\pdfoutput=1 

\usepackage{amsmath,mathtools,amssymb,amsthm,amsxtra,overpic,bbm,epsfig,subfigure,url,bm}
\usepackage{hyperref}
\usepackage{mathrsfs}
\usepackage{color,xcolor}
\usepackage{comment}
\usepackage{float}
\usepackage{enumitem}
\usepackage{slashed}

\newcommand{\prlsection}[2]{{\it\textbf{#1}{#2}}---}
\makeatletter
\newcommand*{\balancecolsandclearpage}{%
	\close@column@grid
	\cleardoublepage
	\twocolumngrid
}
\makeatother

\definecolor{MyDarkBlue}{rgb}{0.1, 0.1, 0.8} 
\definecolor{SBlue}{rgb}{0.2, 0.4, 0.7} 
\definecolor{MyLightBlue}{rgb}{0.22,0.51,0.9}
\definecolor{MyGreen}{rgb}{0.0, 0.5, 0.0}
\definecolor{BrickRed}{rgb}{0.8, 0.25, 0.33}
\definecolor{nicered}{rgb}{0.7,0.12,0.12}
\definecolor{nicegreen}{rgb}{0.,0.5,0.}
\definecolor{niceblue}{rgb}{0.,0.,0.8}
\hypersetup{
	colorlinks=true,
	linkcolor=black,
	filecolor=nicegreen,      
	urlcolor=SBlue,
	citecolor=SBlue,
}

\begin{document}
\title{\bf Probing Cosmic Neutrino Background through Parametric Fluorescence}

\author{Guo-yuan Huang}
\email{huangguoyuan@cug.edu.cn}
\affiliation{School of Mathematics and Physics, China University of Geosciences, 430074 Wuhan, China} 
\author{Shun Zhou}
\email{zhoush@ihep.ac.cn}
\affiliation{Institute of High Energy Physics, Chinese Academy of Sciences, Beijing 100049, China} \affiliation{School of Physical Sciences, University of Chinese Academy of Sciences, Beijing 100049, China} 

\date{\today}
\begin{abstract}
\noindent
We point out that relic neutrinos from the Big Bang may induce the parametric fluorescence in atomic or molecular systems, which offers a novel way to discover cosmic neutrino background. By coherently scattering with molecular energy levels, a massive neutrino can spontaneously ``decay" into a lighter neutrino and an infrared signal photon, i.e., $\nu^{}_{i} + M \to \nu^{}_{j} + \gamma^{}_{\rm S} + M$, where the molecular state $M$ remains unchanged after the scattering. Because the amplitudes of different radiants are matched in phase, the rate is coherently enhanced and proportional to the squared density of ambient dipoles. When the energy transfer from neutrinos coincides with the energy-level difference, the fluorescence will be on resonance. Near the resonance, the rate is proportional to the square of the coherence time $T^{}_{\rm c}$ of the ensemble. For a nominal target volume of $5~{\rm m^3}$ (or $5~{\rm cm^3}$), the signal rate can reach $1~{\rm yr}^{-1}$ for $T^{}_{\rm c} = 10~{\rm ns}$ (or $T^{}_{\rm c} = 10~{\rm \mu s}$). This event rate appears to be very promising in consideration of an even longer coherence time that is achievable in solid systems. 
\end{abstract}

\maketitle

\fontdimen1\font=0.0em
\fontdimen2\font=0.38em
\fontdimen3\font=0.2em
\fontdimen4\font=0.1em

\prlsection{Introduction}{.}The Big Bang theory predicts the existence of both cosmic microwave background (CMB) and cosmic neutrino background (C$\nu$B). While the CMB has been firmly observed and serves as a powerful probe of the early Universe~\cite{Planck:2018vyg}, the C$\nu$B, formed one second after the Big Bang, has yet to be detected. To date, the most promising method for observing these relic neutrinos is through the inverse beta-decay process, initially proposed by S.~Weinberg in 1962~\cite{Weinberg:1962zza}. Nevertheless, the PTOLEMY project, currently under development with tritium targets, still faces  significant challenges before its practical feasibility can be confirmed~\cite{PTOLEMY:2018jst,PTOLEMY:2019hkd,PTOLEMY:2022ldz,Cheipesh:2021fmg,Cheipesh:2023qiy}. An alternative approach to  relic neutrino detection is to observe the mechanical recoils from their coherent scattering on a macroscopic object~\cite{Opher:1974drq,Lewis:1979mu,Zeldovich:1981wf,Cabibbo:1982bb,Shvartsman:1982sn,Langacker:1982ih,Smith:1983jj,Lewis:1987yd,Loeb:1990xs,Ferreras:1995wf,Hagmann:1999kf,Duda:2001hd,Gelmini:2004hg,Ringwald:2009bg,Vogel:2015vfa,Domcke:2017aqj,Shergold:2021evs,Ruzi:2023cvp}. However, while the event rate for coherent scattering can be enhanced by a factor of the Avogadro constant, the resulting  sub-eV recoil is averaged across all the atoms in the target ensemble, posing a fundamental challenge for detecting those elusive relics. For instance, for the mechanical recoil, the resultant acceleration is fifteen orders of magnitude below the most advanced measurements~\cite{Shergold:2021evs}.

\begin{figure}[t!]
	\begin{center}
		\includegraphics[width=0.45\textwidth]{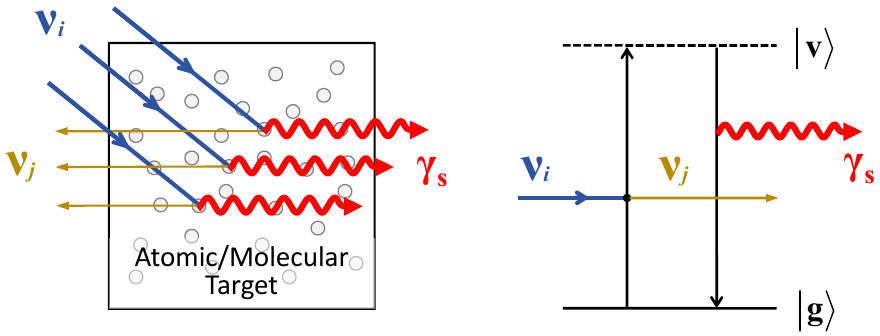}
	\end{center}
	\caption{Schematic diagrams for the parametric fluorescence induced by massive neutrinos, which collectively interact with molecular (or atomic) dipoles in the medium. }
	\label{fig:scheme1}
\end{figure}

In this work, we investigate a novel phenomenon of the coherent parametric fluorescence of relic neutrinos in cold molecular (or atomic) systems. A schematic plot for the process is shown in the left panel of Fig.~\ref{fig:scheme1}. By collectively interacting with molecular dipoles, the parametric fluorescence $\nu^{}_{i} + M \to \nu^{}_{j}+ \gamma^{}_{\rm S} +M$ with neutrino masses $m^{}_{i} > m^{}_{j}$ will be induced, a spontaneous process that is unlikely to occur without the presence of the medium.
Microscopically, neutrinos are scattering with dipoles in the medium through their coupling to electrons, and the process is coherent  for the ensemble because dipoles in the medium remain unmodified after the scattering. 

%
The medium of our concern is characterized by generic electric or magnetic dipoles $\bm{d}$ of Avogadro's number. Those dipoles radiate collectively if phases of their scattering amplitudes are matched. The microscopic process of the parametric fluorescence is described by the diagram on the right panel of Fig.~\ref{fig:scheme1}. We assume that all the molecules in the system are in their ground state $\left|\rm g \right>$, by maintaining a low medium temperature. Neutrino fields couple to the dipole system via the four-fermion interaction, while photons couple to the system via the electric (or magnetic) dipole interaction $\bm{d} \cdot \bm{\mathcal{E}}$ (or $\bm{d} \cdot \bm{\mathcal{B}}$). When a heavier relic neutrino $\nu^{}_{i}$ passes through this medium, it can first excite the dipole to a virtual state $\left|\rm v \right>$ and be converted to a lighter neutrino $\nu^{}_{j}$. The virtual energy level  jumps back to the ground state by radiating a photon $\gamma^{}_{\rm S}$. If the scattering amplitude contributed by each radiant of the medium is represented by $\mathcal{M}^{}_{r}$, then the total amplitude will be
\begin{align} \label{eq:Mtot}
	\mathcal{M}^{}_{\rm tot} = \sum^{N}_{r = 1} \mathcal{M}^{}_{r} \cdot \mathrm{e}^{ \mathrm{i}(\bm{p}^{}_{i}-\bm{p}^{\prime}_{j}- \bm{k}) \cdot \bm{x}^{}_{r} } \;.
\end{align}
Here, $N$ refers to the total number of dipoles in the medium, $\bm{x}^{}_{r}$ is the position of the dipole radiant $\bm{d}^{}_{r}$, and $\bm{p}^{}_{i}$, $\bm{p}^{\prime}_{j}$ and $\bm{k}$ are the momenta of the initial neutrino, the outgoing neutrino and the signal photon, respectively. When the phase matching condition $\bm{q} \equiv \bm{p}^{}_{i}-\bm{p}^{\prime}_{j} - \bm{k} = 0$ is satisfied (for a uniform medium), different dipoles will radiate in phase and the event rate will receive an enhancement of the squared density. 

Since the medium remains unchanged after the fluorescence process, the kinematics for the process $\nu^{}_{i} \to \nu^{}_{j} + \gamma^{}_{\rm S}$ is similar to radiative neutrino decays in vacuum. The energy of the signal photon will be affected by the neutrino mass ordering, which will hopefully be determined within a few years independently by JUNO~\cite{JUNO:2015zny,JUNO:2024jaw} and DUNE~\cite{DUNE:2020lwj,DUNE:2020ypp}.
For concreteness, we take the lightest neutrino mass to be vanishing, and thus the heaviest neutrino mass will be $m^{}_{i} \approx 0.05~{\rm eV}$. In the case of normal mass ordering (NO), the heaviest neutrino will be $\nu^{}_{3}$, and the photon energy for $\nu^{}_3 \to \nu^{}_{1}$ or $\nu^{}_{2}$ should be about $\omega \sim m^{}_{3}/2 \approx 0.025~{\rm eV}$, which falls within the far-infrared range of light. In the case of inverted mass ordering (IO), the transition from the heaviest neutrino will generate a photon with energy about $\omega \sim 0.025~{\rm eV}$ (for $\nu^{}_2 \to \nu^{}_{3}$) or $\omega \sim m^{}_{2}/2 \approx 7.5 \times 10^{-4}~{\rm eV}$ (for $\nu^{}_2 \to \nu^{}_{1}$), the latter of which corresponds to the microwave frequency. While it might be challenging to detect single quantum of microwave, one-photon detection at far-infrared frequencies should be feasible with modern or near-future technologies~\cite{Homma:2024aan}, such as superconducting tunnel junction sensors~\cite{Kim:2018jtj}, kinetic inductance detectors~\cite{Baselmans:2016pdc}, quantum capacitance detectors~\cite{2018NatAs}, etc. 
Therefore, we shall focus on the representative scenario of $\nu^{}_{3}\to \nu^{}_{1}$ in the NO case.

In nonlinear optics, a laser beam $\gamma^{}_{\rm P}$ pumping into certain crystals can be efficiently converted into two photon beams $\gamma^{}_{\rm S}$ and $\gamma^{}_{\rm I}$ at lower energies by coherently scattering with dipoles in the crystal. This process $\gamma^{}_{\rm P} \to \gamma^{}_{\rm S} + \gamma^{}_{\rm I}$ is referred to as the spontaneous parametric down-conversion (SPDC)~\cite{Boyd2008,Klyshko1988,Butcher_Cotter_1990,YanhuaShih2003}, arising from the nonlinear electric susceptibility of the material. 
The conversion efficiency relies largely on detailed  configurations of atoms or molecules, and can be calculated from first principles of quantum mechanics. The most intriguing aspect of this well-established phenomenon is that resonance enhancement occurs when the frequency of the radiation aligns with the corresponding energy-level difference~\cite{Boyd2008,Klyshko1988}. The fluorescence process induced by neutrinos is similar to the pure electromagnetic scenario. This phenomenon should be distinguished from the  soft photon emission due to coherent scattering of neutrinos with free electron gas~\cite{Loeb:1990xs} and very off-shell scatterings with binding electrons for neutrinos at higher energies~\cite{Akhmedov:2018wlf}.

\prlsection{Parametric fluorescence rate for massive neutrinos}{.}The fluorescence processes can be interpreted as the dipole radiation of a polarized medium under the perturbation of external fields. As shown in the right panel of Fig.~\ref{fig:scheme1},
the essence is to have a two-level system with a ground state $\left| \rm g\right>$ and a virtual state $\left| \rm v\right>$. The system is initially prepared in the ground state $\left| \rm g \right>$. Through the virtual state $\left| \rm v\right>$, the scattering process then reads $\nu^{}_{i} + \left| \rm g\right> \to \nu^{}_{j} + \left| \rm g\right> + \gamma$.

The essential way to calculate the fluorescence rate is to determine the  Hamiltonian density for the effective neutrino-photon coupling  induced by dipoles.
In nonlinear optics, there is a standard procedure to extract the Hamiltonian density for the electromagnetic couplings~\cite{Boyd2008,Klyshko1988,Butcher_Cotter_1990}, by calculating the electric or magnetic polarization (i.e., the induced dipole moment per volume) $\bm{P}$ in response to the field perturbation. 
The polarization  reflects the strength of the medium optical response when an external field is applied.
The Hamiltonian density describing this response can be straightforwardly obtained with $\mathcal{H}^{}_{\rm eff} = -\bm{P} \cdot \bm{\mathcal{E}}$ or $\mathcal{H}^{}_{\rm eff} = -\bm{P} \cdot \bm{\mathcal{B}}$ depending on whether this polarization is electric or magnetic.
For reference, we briefly outline the procedure for the pure electromagnetic processes in the Supplemental Material. The effective Hamiltonian density and transition matrix for neutrino interactions can be derived in a similar manner.

Neutrino current can induce both the M1 and E1 transitions; see the Supplemental Material for more details.
In our context, the M1 transition is preferred because its neutrino-induced current is of order one, while the corresponding current in the E1 case is suppressed by the Bohr radius. 
Hence, in the following we should focus on the M1 transition, and the result for the E1 case will also be presented in the Supplemental Material.
In general, the quantum state of the system can be described as a superposition of discrete energy levels,
\begin{align}
	\left| \Psi(t) \right> = C^{}_{\rm g}(t) \left| \rm g \right> + \sum^{}_{\rm v}C^{}_{\rm v}(t) \left| \rm v \right> \;,
\end{align}
where $C^{}_{\rm g}$ and $C^{}_{\rm v}$ are coefficients of the ground state and various excited states, respectively.
The initial state is set to be $\left| \Psi^{}_{0} \right> = \left| \rm g \right>$.

For a two-level system connected via the M1 transition, the relevant Hamiltonian describing the neutrino and photon interactions is represented by Eq.~(9) of the Supplemental Material.
In this case, the M1 transition corresponds to a flip in the electron spin. 
Those two spin configurations can feature different energies in the presence of couplings to the orbital angular momentum or a static magnetic field.
The coefficient for $\left| \rm v \right>$ under the perturbation of neutrino fields reads
\begin{align}
	C^{(1)}_{\rm v} (t) =   & \frac{G^{}_{\rm F} C^{}_{ij} }{\sqrt{2}} 
	\frac{\hat{\bm{j}}^{}_{i j} \cdot \bm{\sigma}^{}_{\rm vg} }{E^{}_{\rm vg} + E^{\prime}_{j} - E^{}_{i}} \mathrm{e}^{\mathrm{i}(E^{}_{\rm vg} + E^{\prime}_{j} - E^{}_{i}) t}\;,
\end{align}
where $G^{}_{\rm F}$ is the Fermi constant and $C^{}_{ij} \equiv U^*_{ei} U^{}_{ej}$ ($i\neq j$) with $U$ being the Pontecorvo-Maki-Nakagawa-Sakata (PMNS) matrix, $E^{}_{\rm vg}$ is the energy difference between $\left| \rm v \right>$ and $\left| \rm g \right>$, $\hat{\bm{j}}^{}_{i j}$ denotes the neutrino current operator and $\bm{\sigma}^{}_{\rm vg}\equiv \left< \rm v\right| \bm{\sigma} \left| \rm g \right>$ with $\bm{\sigma}$ being the Pauli matrices. The polarization can be obtained similar to the pure electromagnetic case, and we present the resultant effective Hamiltonian density below, 
\begin{align} \label{eq:HnngM}
	\mathcal{H}^{\rm M1}_{\nu\nu\gamma} =   & -\frac{G^{}_{\rm F} C^{}_{ij} }{\sqrt{2} } 
	\frac{n^{}_{{d}}\, [ \hat{\bm{j}}^{}_{i j} \cdot \bm{\sigma}^{}_{\rm vg} ]\, [\bm{d}^{*}_{\rm vg} \cdot \bm{\mathcal{B}}(\bm{k})] }{E^{}_{\rm vg} + E^{\prime}_{j} - E^{}_{i}} \notag \\
	& \times \mathrm{e}^{\mathrm{i}(\omega + E^{\prime}_{j} - E^{}_{i} ) t }  +  {\rm h.c.}\;,
\end{align}
where $n^{}_{{d}} \sim N^{}_{\rm A}~{\rm cm^{-3}}$ represents the number density of atomic/molecular dipoles in a solid with $N^{}_{\rm A} \approx 6.02 \times 10^{23}$ being the Avogadro constant.
The temporal phase  explicitly shown implies the energy conservation.
Other than the temporal phase, $\mathcal{H}^{\rm M1}_{\nu\nu\gamma}$ also carries a spatial phase in  field operators, namely
$\mathrm{exp}[\mathrm{i}(\bm{p}^{}_{i}-\bm{p}^{\prime}_{j}- \bm{k}^{}_{}) \cdot \bm{x}^{}_{r}]$ as in Eq.~(\ref{eq:Mtot}).
For a uniform medium, after integrating over the target volume
this phase will result in the delta function $\delta^{(3)}_{}(\bm{p}^{}_{i}-\bm{p}^{\prime}_{j}- \bm{k}^{}_{})$, which imposes the phase-matching condition (identical to the momentum conservation in vacuum) for the process to be coherent. 

The transition matrix for $\nu^{}_{i} \to \nu^{}_{j}\gamma$ is obtained by an integration $\mathrm{i}T = -\mathrm{i}\int^{\infty}_{-\infty} \mathrm{d}^4 x \big<\bm{p}^{\prime}_{j}, \bm{k} \big| \mathcal{H}^{}_{\nu\nu\gamma} \big| \bm{p}^{}_{i} \big>$.
For the uniform material, the kinematics of the transition is analogous to the case of the decay of non-relativistic neutrinos in vacuum, and the decay products $\nu^{}_{j}$ and $\gamma$ will be back-to-back.
Following the standard formula, the differential decay rate in the rest frame of $\nu^{}_{i}$ with respect to the photon outgoing direction is 
\begin{align} \label{eq:dGdO}
	\frac{\mathrm{d} \Gamma}{\mathrm{d} \Omega^{}_{\gamma}} = \int  \mathrm{d} |\bm{k}|\frac{ \bm{k}^2\, |\mathcal{M}|^2}{32 \pi^2 m^{}_{i} E^{\prime}_{j}\omega } \delta (\omega + E^{\prime}_{j} - E^{}_{i})\;.
\end{align}
The integration over the delta function will generate a dimensionless factor 
$D^{}_{k} \equiv 1/[\mathrm{d}(\omega + E^{\prime}_{j} - E^{}_{i})/\mathrm{d}|\bm{k}|]$.
The actual value of $D^{}_{k}$ relies on  neutrino masses as well as the dispersion relation of photons in medium.
For $\nu^{}_{3} \to \nu^{}_{1}$ with a nearly massless $\nu^{}_{1}$  in the NO case, the energy of daughter particles just takes $E^{\prime}_{1} = \omega = m^{}_{3}/2$, assuming the photon dispersion relation in vacuum. The value of $D^{}_{k}$ is just 2 in this case.
The in-medium effect causes the energy of the signal photon to deviate from $m^{}_{3}/2$, which will be discussed later. 

We explicitly perform the angular integration, arriving at the  rate
\begin{align}\label{eq:Gamma}
	\Gamma & \approx  \frac{G^2_{\rm F} |U^*_{e i} U^{}_{e j}|^2 }{2 \pi } \frac{f^{\rm M1}_{\theta}\, n^2_{\bm{d}}\, |\bm{d}^{}_{\rm vg}|^2 D^{}_{k}}{(E^{}_{\rm vg} + E^{\prime}_{j} - E^{}_{i})^2 + \gamma^2_{\rm vg}/4} \frac{ |\bm{k}|^4 } {\omega}\;,  
\end{align}
where $f^{\rm M1}_{\theta}\sim \mathcal{O}(1)$ is the form factor depending on the polarization of magnetic dipoles. The outgoing neutrino is taken to be relativistic in Eq.~(\ref{eq:Gamma}). 
The damping rate $\gamma^{}_{\rm vg} \equiv T^{-1}_{\rm c}$ accounts for all factors that can affect the timescale of coherent accumulation of $\left|\rm v \right>$, where $T^{}_{\rm c}$ denotes the overall coherent timescale. 
When $E^{}_{\rm vg} = E^{}_{i} - E^{\prime}_{j}$ is satisfied, the transition will be on resonance, and the rate will approximately be $\Gamma \sim G^{2}_{\rm F} n^{2}_{\bm{d}} |\bm{d}^{}_{\rm vg}|^2 |\bm{k}|^3 T^{2}_{\rm c}$ with $|\bm{d}^{}_{\rm vg}| \sim e/m^{}_{e}$ (i.e., the Bohr magneton) being the magnetic dipole moment. 
We  notice that $T^{}_{\rm c}$ is a key parameter for our rate estimation. The coherence time is mainly contributed by  the spontaneous decay lifetime $T^{}_1$ of $\left|\rm v \right>$ and the pure dephasing time $T^{}_{\phi}$, according to the convention $T^{-1}_{\rm c} = T^{-1}_1 + T^{-1}_{\phi}$~\cite{fleischhauer2005EIT}. For energy splittings of $\mathcal{O}(25~{\rm meV})$, the decay lifetime $T^{}_{1} \sim 1/(d^{2}_{\rm vg} E^{3}_{\rm vg})$ is relatively long (e.g., $120~{\rm s}$ for M1 and $6~{\rm ms}$ for E1 transitions), implying that coherence is mainly limited by the dephasing time $T^{}_{\phi}$ arising from stochastic dipole-environment interactions. Over the past decades, substantial AMO advances  have demonstrated spin coherence times  in solids controllable at the $\mathcal{O}(10~{\rm ms})$ level and even longer~\cite{Fraval2004,bar2013solid,herbschleb2019ultra,upadhyay2020ultralong,kuznetsova2002atomic,caldwell2020long,heinze2013stopped}, 
highlighting the potential for achieving long-lived coherence in such systems.

The resonance enhancement also suggests that molecular materials are ideal target candidates, as their energy-level differences typically fall within the infrared range.
In an ideal case that the incoming neutrino flux with a density $n^{}_{\nu_i} \approx 112~{\rm cm^{-3}}$ (including both neutrinos and antineutrinos) can satisfy the resonance condition, the event rate $R = n_{\nu_i}\, V \, \Gamma$ in a material target of volume $V$ will roughly be given by 
\begin{align} \label{eq:R31M1A}
	R^{}_{\nu_3 \to \nu_1} &\sim 1~{\rm yr^{-1}} \hspace{0.3cm}(V=5~{\rm m}^3,\, T^{}_{\rm c} = 10~{\rm ns}) \;, \\\label{eq:R31M1B}
	R^{}_{\nu_3 \to \nu_1} &\sim 8~{\rm yr^{-1}} \hspace{0.3cm}(V=40~{\rm cm}^3,\, T^{}_{\rm c}=10~{\rm \mu s}) \;.
\end{align}
assuming a resonance energy of $E^{}_{\rm vg} = 0.025~{\rm eV}$.
Due to its $T^2_{\rm c}$ dependence, the event rate appears highly promising if a long coherence time can be achieved, and a coherence time of $10~{\rm \mu s}$ should be well achievable with existing techniques~\cite{kuznetsova2002atomic,Fraval2004,bar2013solid,herbschleb2019ultra,upadhyay2020ultralong,kuznetsova2002atomic}, particularly at low temperatures. Recall that in the case of Majorana neutrinos, the overall event rate will be doubled.

A  condition for achieving the resonance rate in Eqs.~(\ref{eq:R31M1A}) and (\ref{eq:R31M1B}) seems to have a highly monoenergetic neutrino flux that matches the energy level. Relic neutrinos with a temperature $1.95~{\rm K}$ in the Milky Way exhibit a momentum dispersion of order $\Delta p \sim 10^{-4}~{\rm eV}$, leading to a tiny energy spread of order $\Delta E^{}_{3} \approx \Delta p^2/(2m^{}_{3}) \sim 10^{-6}~{\rm eV}$ for the incoming non-relativistic neutrinos. The corresponding timescale of the energy spread is on the order of $1~{\rm ns}$. Nevertheless, the actual resonance condition $E^{}_{\rm vg} = E^{}_{3} - E^{\prime}_{1}$ imposes stricter requirements, even if $E^{}_{3}$ is already quite monoenergetic.
Because the outgoing neutrino and photon are both ultra-relativistic for $\nu^{}_{3} \to \nu^{}_{1} \gamma$ with $E^{\prime}_{1} \approx |\bm{p}^{\prime}_{1}|$ and $\omega = k$, this momentum spread actually results in a spread of comparable magnitude $\Delta(E^{}_{3} - E^{\prime}_{1}) \sim 10^{-4}~{\rm eV}$. This energy spread will limit the resonance enhancement, because only a small portion of the neutrino flux ($\sim \gamma^{}_{\rm vg} / \Delta p$ for $\Delta p \gg \gamma^{}_{\rm vg}$) can cover the resonance peak.
Thus, the average decay rate seems to be capped at 
$\Gamma \sim G^{2}_{\rm F} n^{2}_{\bm{d}} |\bm{d}^{}_{\rm vg}|^2 \omega^3 / (\gamma^{}_{\rm vg}\Delta p )$. In this case, to reach the event rate of $1~{\rm yr}^{-1}$, a coherence timescale of $4~{\rm ms}$ is required for the target volume $V=40~{\rm cm}^3$.

The discussions so far have not accounted for in-medium effects, which can alter photon's dispersion relation, thus affecting our rate estimates. 
This is unavoidable for applying our results to realistic detection, as resonance, which is our region of interest, typically strongly modifies the refractive index.
Remarkably, we find that including the in-medium effect is not harmful and can even help by reducing the impact of relic neutrinos' momentum spread.
The key is the slow-light phenomenon well recognized in modern quantum optics~\cite{hau1999light,Khurgin2009,Khurgin2010}, which can render the outgoing photon effectively non-relativistic with a reduced group velocity as $\mathcal{O}(10)~{\rm m/s}$ or even lower~\cite{turukhin2001observation,kuznetsova2002atomic,schraft2016stopped}.
The group velocity is connected to photon's dispersion curve through $v^{}_{\rm g} = \mathrm{d}\omega/\mathrm{d}{k}$.
With this effect, we find that the energy spread can be controlled within $\Delta(E^{}_{3} - E^{\prime}_{1}) \sim (10^{-4} \cdot v^{}_{\rm g})~{\rm eV}$. 

Interestingly, the slow light intrinsically occurs in the vicinity of the resonance~\cite{Khurgin2010}. This is unsurprising, as most slow-light schemes rely on photon scattering with some resonant structure of the medium, either discrete energy levels or photonic resonators~\cite{Khurgin2009}. The reduced group velocity arises from the resonant energy transfer between photons and other modes, such as the polarization of the non-relativistic medium.

\begin{figure}[t!]
	\begin{center}
		\includegraphics[width=0.4\textwidth]{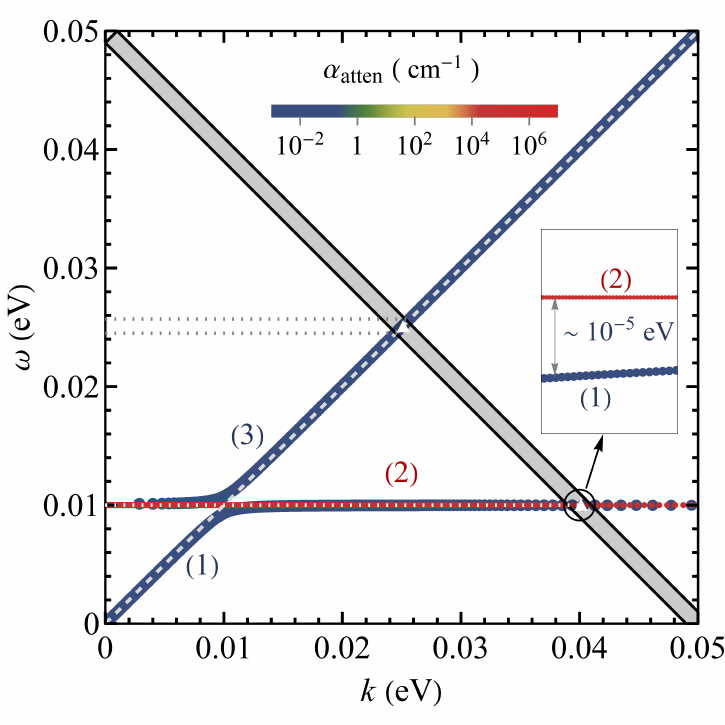} 
	\end{center}
	\caption{The dependence curves of the photon energy $\omega$ (colorful curves) and the energy difference of neutrinos $E^{}_{3} - E^{\prime}_{1}$ (gray band) on the photon momentum $k$, where $m^{}_{3}=0.05~{\rm eV}$ and $m^{}_{1}=0~{\rm eV}$ are inputs. The shaded gray region denotes a momentum spread of order $10^{-3}~{\rm eV}$ for illustration. The intersection points marked by white triangles represent the resultant momentum and energy satisfying the conservation law. For photon's dispersion relation, the following parameters have been used for demonstration: $E^{}_{\rm vg} = 10~{\rm meV}$, $d^{}_{\rm vg} = e/(4 m^{}_{e})$, $n^{}_{d} = 6.02 \times 10^{23}~{\mathrm{cm}}^{-3}$ and $T^{}_{\rm c} = 10~{\rm \mu s}$. Similar dispersion curves can be found in Ref.~\cite{Khurgin2010}.
	}
	\label{fig:dispersion}
\end{figure}

In Fig.~\ref{fig:dispersion}, we show the dispersion curve for a two-level system, with detailed parameters provided in the caption. For illustration, we consider an energy level separation of $E^{}_{\rm vg} = 10~{\rm meV}$. Near the resonance energy, the dispersion splits into three branches with noticeable flattening, a well known feature of slow-light schemes~\cite{Khurgin2010}. We find a group velocity as low as $10^{-8}\,c$ for our chosen parameters. For more details, we refer the readers to the Supplemental Material.

In practice, incoherent absorption attenuates the signal fluorescence during its propagation in the medium. This dissipative process does not imply the complete loss of energy, as it could eventually be down-converted into several lower-frequency photons and optical phonons. 
Nevertheless, such incoherent effects limit the number of ambient radiants available for neutrino coherent scatterings, which require a propagation length at least comparable to the de Broglie wavelength of the outgoing photon, i.e., $l^{}_{\rm atten}  > 1/k^{}_{} \sim 5~{\rm \mu m}$.
Without additional mechanisms, the attenuation length (1/$\alpha^{}_{\rm atten}$) at the resonance peak (e.g., in the second branch in Fig.~\ref{fig:dispersion}) for the given example  is $0.03~{\rm \mu m}$, limiting the coherence application. 
Slight detuning from the exact resonance can significantly improve the attenuation, though at a cost of weaker resonance enhancement.
In Fig.~\ref{fig:dispersion}, the first branch near $k = 0.04~{\rm eV}$ corresponds to a detuning of $\Delta^{}_{\rm p} \equiv  \omega -E^{}_{\rm vg} \approx -10^{-5}~{\rm eV}$, leading to a  small rate.
However, we find that decreasing the dipole strength can shift this branch much closer to the resonance peak. By adopting $d^{}_{\rm vg} = e/(250\, m^{}_{e})$, we obtain a comparable rate to Eq.~(\ref{eq:R31M1A}), with a detuning $-1.7 \times 10^{-9}~{\rm eV}$, attenuation length $l^{}_{\rm atten} \approx 0.3~{\rm mm}$ and $v^{}_{\rm g} \approx 10^{-7}\,c$. For $d^{}_{\rm vg} = e/(10^4\, m^{}_{e})$ and a reachable coherence $T^{}_{\rm c} = 10~{\rm ms}$, we obtain $R \sim 1~{\rm yr^{-1}}$ with $V=10^{-3}~{\rm m^3}$, $ \Delta^{}_{\rm p} \approx -10^{-12}~{\rm eV}$,  $l^{}_{\rm atten} \approx 0.3~{\rm mm}$ and $v^{}_{\rm g} \approx 10^{-10}\, c$. Further decreasing the dipole strength will loose the flattening feature up to $k = 0.04~{\rm eV}$.

\begin{figure}[b!]
	\begin{center}
		\includegraphics[width=0.47\textwidth]{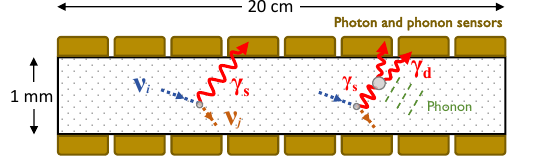}
	\end{center}
	\caption{A schematic plot of the experimental setup for detecting parameter fluorescence. The target is a solid film with dimensions $20~{\rm cm} \times 20~{\rm cm} \times 1~{\rm mm}$. Its surface is covered with superconducting photon and phonon sensors featuring $\mathcal{O}(10~{\rm meV})$ sensitivities. Two types of signals can arise, depending on whether the initial photon is absorbed by the medium before reaching the sensors. Similar geometry may be found in experiments searching for light dark matters~\cite{Temples:2024ntv,Temples:2025xew}.}
	\label{fig:scheme}
\end{figure}

It is also recognized that attenuation can be suppressed even at exact resonance through the electromagnetically induced transparency (EIT), a powerful technique widely used in slow-light experiments~\cite{fleischhauer2005EIT,harris1997electromagnetically,marangos1998electromagnetically,lukin2001controlling,finkelstein2023practical,kuznetsova2002atomic}.
As shown in the Supplemental Material, by irradiating electromagnetic waves to upper hyperfine sub-levels, e.g., separated by microwave frequencies, the medium can become transparent (with  attenuation length $l^{}_{\rm atten}  > 1~{\rm mm}$) to photons at the complete resonance frequency.
We have not explored this effect in full detail, but a potential rate in Eq.~(\ref{eq:R31M1B}) without the attenuation concern is fairly anticipated.

The target can be designed in a disc-like shape with a thickness on the $\mathcal{O}({\rm mm})$ scale. A schematic of the experimental  setup is shown in Fig.~\ref{fig:scheme}, where the target dimensions can be $20~{\rm cm} \times 20~{\rm cm} \times 1~{\rm mm}$. 
The target surface is covered with superconducting sensors sensitive to $\mathcal{O}(10)~{\rm meV}$ photons or phonons~\cite{golwala2022novel,Essig:2022dfa}. A variety of sensing technologies targeting for a few meV energy deposition are  under active developments, primarily driven by the light dark matter detection~\cite{Lyon:2022sza,Cruciani:2022mbb,Gao:2024irf,Temples:2024ntv,Ramanathan:2024hsf,Baudis:2025zyn,TESSERACT:2025tfw,Sandoval:2025mye,Temples:2025xew}.
This configuration also lends itself to stacking multiple layers. 
To minimize backgrounds, this device should be installed deep underground; see Supplemental Materials for discussions about possible backgrounds.
If signal photons reach the material surface, they can break Cooper pairs in the sensor's absorber, generating readout signals. During propagation, these photons may further interact with the material, down-converted into secondary photons and phonons, which are also detectable. Nonetheless, this requires dedicated simulations given a specific material.

\prlsection{Summary and discussion}{.}In analogy to nonlinear optics, neutrinos can experience the parametric fluorescence in  material targets with a photon spontaneously emitted. This process is governed by the standard weak interaction of neutrinos and does not rely on any electromagnetic couplings of neutrinos in vacuum.
The basic condition is no more than the presence of clean bulk material at low temperatures.
The signal is an infrared photon from the dark medium environment.
By coherently interacting with this material, neutrinos can develop a rather strong effective coupling to the electromagnetic field, e.g., Eq.~(\ref{eq:HnngM}). The strength of this coupling depends largely on the detailed configurations of atomic or molecular energy levels of the material. A notable scenario is the resonance fluorescence, which occurs when the energy transferred by the neutrino matches the dipole frequency. The event rate of $\nu^{}_{3}\to \nu^{}_{1}\gamma$ on resonance is  considerable for the M1 transition. In the  conservative case with the coherence time $T^{}_{\rm c} = 10~{\rm ns}$, we have $R\sim 1~{\rm yr^{-1}}$ for a target volume $V = 5~{\rm m^3}$. In an achievable case with $T^{}_{\rm c} = 10~{\rm \mu s}$, we only need $V = 40~{\rm {cm}^3}$ to have a rate $R\sim 8~{\rm yr^{-1}}$.

{We further remark on additional possible enhancement of the event rate, even though prolonging the coherence timescale already represents a promising direction.}
The transition rate under consideration is governed by the magnitudes of the neutrino and electron currents. In the case of the M1 transition, the neutrino current is of order one, while the electron current is suppressed by a factor of 
$\alpha^2 \sim 5 \times 10^{-5}$. Conversely, for the E1 transition, the suppression factor $m^{2}_{i} a^2_0 \sim 10^{-10}$ appears in the neutrino current instead while the electron current is not suppressed.
%
%
This indicates configuration mixing as a potential opportunity for significant enhancement through configuration mixing between opposite parities~\cite{Bransden2006}. Parity mixing renders the transition to simultaneously proceed  through both E1 and M1 channels, thus evading the suppression that individually affects each mode. Configuration mixing  can be achieved with many methods, such as the Stark effect in the presence of an external electric field. The Stark effect has been extensively employed in experiments probing atomic parity violation~\cite{Bouchiat:1975zz,BOUCHIAT1982358}, owing to its ability to induce sizable parity mixing.
{For a full parity mixing~\cite{gong2024electric}, we could have an enhancement factor up to $\sim 10^{4}$ compared to the pure M1 case.}
	%
Another possible consideration is using the superradiance of a $\Lambda$-like system, where the successive transitions are of M1 and E1 types, respectively, with the neutrino coupling to the first transition and the photon coupling to the second. 
The superradiance mechanism itself has already been considered as an enhancement method to probe neutrino physics~\cite{Fukumi:2012rn,Yoshimura:2012tm,Dinh:2012qb,Song:2015xaa,Zhang:2016lqp,Huang:2019phr,Ge:2021lur,Ge:2022cib,Ge:2023oag}, including the absolute scale of neutrino masses first proposed in Ref.~\cite{Yoshimura:2006nd} as well as relic neutrinos~\cite{Yoshimura:2014hfa,Arvanitaki:2024taq}.

The detection of relic neutrinos can also help to pin down basic neutrino properties, such as the absolute scale of neutrino masses and the Majorana nature. Besides relic neutrinos, the parametric fluorescence mechanism explored in the present work may also be applied to the atomic detection of dark matter~\cite{Safronova:2017xyt,Arvanitaki:2017nhi,Huang:2019rmc,Bhoonah:2019eyo,Bauer:2024dbg} and other monoenergetic neutrino fluxes such as M\"ossbauer neutrinos. 
{To make the detection concept feasible, several challenges must be addressed, including identifying suitable materials among a broad range of atoms or molecules, coherence time controlling, infrared photon or phonon sensing, and background reduction. Achieving this goal will call for joint efforts and expertise from both the particle and AMO physics communities.}
Since the main purpose of this work is to put forward a promising direction for relic neutrino detection, its ultimate experimental realization will be addressed in future works.

\prlsection{Acknowledgments}{.}{\it The authors would like to thank Wei Chao, Christina Gao and Shao-Feng Ge for helpful discussions and comments. This work is supported in part by the ``CUG Scholar'' Scientific Research Funds at China University of Geosciences (Wuhan) under project No. 2024014, and by the National Natural Science Foundation of China under grant No. 12475113 and grant No. 12405130.}

\bibliographystyle{utcaps_mod}
\bibliography{reference}

@article{gong2024electric,
	title={Electric-field-induced mixing of opposite parity states: Insights from spectra of the CO (b 3 $\Sigma$+← a 3 $\Pi$ $\Omega$) transition},
	author={Gong, Shiyan and Wang, Peng and Mo, Yuxiang},
	journal={Physical Review A},
	volume={110},
	number={5}, 
	pages={052801},
	year={2024},
	doi = "10.1103/PhysRevA.110.052801",
	publisher={APS}
}

@article{Fraval2004,
	title = {Method of Extending Hyperfine Coherence Times in ${\mathrm{P}\mathrm{r}}^{3+}\ensuremath{\mathbin:}{\mathrm{Y}}_{2}{\mathrm{S}\mathrm{i}\mathrm{O}}_{5}$},
	author = {Fraval, E. and Sellars, M. J. and Longdell, J. J.},
	journal = {Phys. Rev. Lett.},
	volume = {92},
	issue = {7}, 
	pages = {077601},
	numpages = {4},
	year = {2004},
	doi = "10.1103/PhysRevLett.92.077601",
	publisher = {American Physical Society}
}

@article{herbschleb2019ultra,
	title={Ultra-long coherence times amongst room-temperature solid-state spins},
	author={Herbschleb, ED and Kato, H and Maruyama, Y and Danjo, T and Makino, T and Yamasaki, S and Ohki, I and Hayashi, K and Morishita, H and Fujiwara, M and others},
	journal={Nature communications},
	volume={10},
	number={1},
	pages={3766},
	year={2019},
	doi = "10.1038/s41467-019-11776-8",
	publisher={Nature Publishing Group UK London}
}

@article{bar2013solid,
	title={Solid-state electronic spin coherence time approaching one second},
	author={Bar-Gill, Nir and Pham, Linh M and Jarmola, Andrejs and Budker, Dmitry and Walsworth, Ronald L},
	journal={Nature communications},
	volume={4},
	number={1},
	pages={1743},
	year={2013},
	doi = "10.1038/ncomms2771",
	publisher={Nature Publishing Group UK London}
}

@article{TESSERACT:2025tfw,
	author = "Bui, T. K. and others",
	collaboration = "TESSERACT",
	title = "{First Limits on Light Dark Matter Interactions in a Low Threshold Two-Channel Athermal Phonon Detector from the TESSERACT Collaboration}",
	eprint = "2503.03683", 
	archivePrefix = "arXiv",
	primaryClass = "hep-ex",
	doi = "10.1103/hsrl-crvf",
	journal = "Phys. Rev. Lett.",
	volume = "135",
	number = "16",
	pages = "161002",
	year = "2025"
}

@article{Baxter:2022dkm,
	author = "Baxter, Daniel and others",
	title = "{Snowmass2021 Cosmic Frontier White Paper: Calibrations and backgrounds for dark matter direct detection}",
	eprint = "2203.07623",
	archivePrefix = "arXiv",
	primaryClass = "hep-ex",
	reportNumber = "FERMILAB-PUB-22-205-ND-PPD-QIS",
	month = "3",
	year = "2022"
}

@article{Baryakhtar:2018doz,
	author = "Baryakhtar, Masha and Huang, Junwu and Lasenby, Robert",
	title = "{Axion and hidden photon dark matter detection with multilayer optical haloscopes}",
	eprint = "1803.11455",
	archivePrefix = "arXiv",
	primaryClass = "hep-ph",
	doi = "10.1103/PhysRevD.98.035006",
	journal = "Phys. Rev. D",
	volume = "98",
	number = "3",
	pages = "035006",
	year = "2018"
}

@article{Du:2020ldo,
	author = "Du, Peizhi and Egana-Ugrinovic, Daniel and Essig, Rouven and Sholapurkar, Mukul",
	title = "{Sources of Low-Energy Events in Low-Threshold Dark-Matter and Neutrino Detectors}",
	eprint = "2011.13939",
	archivePrefix = "arXiv",
	primaryClass = "hep-ph",
	doi = "10.1103/PhysRevX.12.011009",
	journal = "Phys. Rev. X",
	volume = "12",
	number = "1",
	pages = "011009",
	year = "2022"
}

@article{Derenzo:2018plr,
	author = "Derenzo, S. and Bourret, E. and Hanrahan, S. and Bizarri, G.",
	title = "{Cryogenic Scintillation Properties of $n$-Type GaAs for the Direct Detection of MeV/c$^2$ Dark Matter}",
	eprint = "1802.09171",
	archivePrefix = "arXiv",
	primaryClass = "physics.ins-det",
	doi = "10.1063/1.5018343",
	journal = "J. Appl. Phys.",
	volume = "123",
	number = "11",
	pages = "114501",
	year = "2018"
}

@article{Derenzo:2016fse,
	author = "Derenzo, Stephen and Essig, Rouven and Massari, Andrea and Soto, Adr{\'\i}an and Yu, Tien-Tien",
	title = "{Direct Detection of sub-GeV Dark Matter with Scintillating Targets}",
	eprint = "1607.01009",
	archivePrefix = "arXiv",
	primaryClass = "hep-ph",
	doi = "10.1103/PhysRevD.96.016026",
	journal = "Phys. Rev. D",
	volume = "96",
	number = "1",
	pages = "016026",
	year = "2017"
}

@article{Cruciani:2022mbb,
	author = "Cruciani, A. and others",
	title = "{BULLKID: Monolithic array of particle absorbers sensed by kinetic inductance detectors}",
	eprint = "2209.14806",
	archivePrefix = "arXiv",
	primaryClass = "physics.ins-det",
	doi = "10.1063/5.0128723",
	journal = "Appl. Phys. Lett.",
	volume = "121",
	number = "21",
	pages = "213504", 
	year = "2022"   
}

@inproceedings{Temples:2025xew,
	author = "Temples, Dylan J. and others",
	title = "{Development Status of the KIPM Detector Consortium}",
	eprint = "2509.25544",
	archivePrefix = "arXiv",
	primaryClass = "physics.ins-det",
	reportNumber = "FERMILAB-PUB-25-0652-ETD",
	month = "9",
	year = "2025"
}

@inproceedings{Essig:2022dfa,
	author = "Essig, Rouven and others",
	title = "{Snowmass2021 Cosmic Frontier: The landscape of low-threshold dark matter direct detection in the next decade}",
	booktitle = "{Snowmass 2021}",
	eprint = "2203.08297",
	archivePrefix = "arXiv",
	primaryClass = "hep-ph",
	reportNumber = "FERMILAB-CONF-22-181-PPD",
	month = "3",
	year = "2022"
}

@article{Ramanathan:2024hsf,
	author = "Ramanathan, Karthik and Parker, John E. and Joshi, Lalit M. and Beyer, Andrew D. and Echternach, Pierre M. and Rosenblum, Serge and Sandoval, Brandon J. and Golwala, Sunil R.",
	title = "{Quantum Parity Detectors: a qubit based particle detection scheme with meV thresholds for rare-event searches}",
	eprint = "2405.17192",
	archivePrefix = "arXiv",
	primaryClass = "physics.ins-det",
	month = "5",
	year = "2024"
}

@inproceedings{Sandoval:2025mye,
author = "Sandoval, Brandon J. and Beyer, Andrew D. and Echternach, Pierre M. and Golwala, Sunil R. and Ho, William D. and Yuan, Lanqing and Ramanathan, Karthik",
title = "{Assessing the operating characteristics of an ion-milled phonon-mediated quantum parity detector}",
eprint = "2509.18637",
archivePrefix = "arXiv",
primaryClass = "physics.ins-det",
month = "9",
year = "2025"
}

@article{Temples:2024ntv,
	author = "Temples, Dylan J. and others",
	title = "{Performance of a phonon-mediated kinetic inductance detector at the NEXUS cryogenic facility}",
	eprint = "2402.04473",
	archivePrefix = "arXiv",
	primaryClass = "physics.ins-det",
	reportNumber = "FERMILAB-PUB-23-674-LDRD-PPD",
	doi = "10.1103/PhysRevApplied.22.044045",
	journal = "Phys. Rev. Applied",
	volume = "22",
	number = "4",
	pages = "044045",
	year = "2024"
}

@article{golwala2022novel,
	title={Novel quantum sensors for light dark matter and neutrino detection},
	author={Golwala, Sunil R and Figueroa-Feliciano, Enectali},
	doi = "10.1146/annurev-nucl-102020-112133",
	journal={Annual Review of Nuclear and Particle Science},
	volume={72},
	pages={419--446},
	year={2022},
	publisher={Annual Reviews}
}

@article{Baudis:2025zyn,
	author = "Baudis, Laura and others",
	title = "{First Sub-MeV Dark Matter Search with the QROCODILE Experiment Using Superconducting Nanowire Single-Photon Detectors}",
	doi = "10.1103/4hb6-f6jl",
	journal = "Phys. Rev. Lett.",
	volume = "135",
	number = "8",
	pages = "081002",
	year = "2025"
}

@article{Gao:2024irf,
	author = "Gao, Jiansong and Hochberg, Yonit and Lehmann, Benjamin V. and Nam, Sae Woo and Szypryt, Paul and Vissers, Michael R. and Xu, Tao",
	title = "{Detecting Light Dark Matter with Kinetic Inductance Detectors}",
	eprint = "2403.19739",
	archivePrefix = "arXiv",
	primaryClass = "hep-ph",
	reportNumber = "MIT-CTP/5654",
	month = "3",
	year = "2024"
}

@article{Lyon:2022sza,
	author = "Lyon, S. A. and Castoria, Kyle and Kleinbaum, Ethan and Qin, Zhihao and Persaud, Arun and Schenkel, Thomas and Zurek, Kathryn M.",
	title = "{Single phonon detection for dark matter via quantum evaporation and sensing of He3}",
	eprint = "2201.00738",
	archivePrefix = "arXiv",
	primaryClass = "hep-ex",
	doi = "10.1103/PhysRevD.109.023010",
	journal = "Phys. Rev. D",
	volume = "109",
	number = "2",
	pages = "023010",
	year = "2024"
}

@article{heinze2013stopped,
	title={Stopped Light and Image Storage by Electromagnetically Induced Transparency up to the Regime of One Minute},
	author={Heinze, Georg and Hubrich, Christian and Halfmann, Thomas},
	journal={Physical review letters},
	volume={111},
	number={3},
	pages={033601},
	year={2013},
	publisher={APS}
}

@article{schraft2016stopped,
	title={Stopped light at high storage efficiency in a Pr 3+: Y 2 SiO 5 crystal},
	author={Schraft, Daniel and Hain, Marcel and Lorenz, Nikolaus and Halfmann, Thomas},
	journal={Physical review letters},
	volume={116},
	number={7},
	pages={073602},
	year={2016},
	publisher={APS}
}

@article{turukhin2001observation,
	title={Observation of ultraslow and stored light pulses in a solid},
	author={Turukhin, AV and Sudarshanam, VS and Shahriar, MS and Musser, JA and Ham, BS and Hemmer, PR},
	journal={Physical Review Letters},
	volume={88},
	number={2},
	pages={023602},
	year={2001},
	publisher={APS}
}

@article{finkelstein2023practical,
	title={A practical guide to electromagnetically induced transparency in atomic vapor},
	author={Finkelstein, Ran and Bali, Samir and Firstenberg, Ofer and Novikova, Irina},
	journal={New Journal of Physics},
	volume={25},
	number={3},
	pages={035001},
	year={2023},
	publisher={IOP Publishing}
}

@article{upadhyay2020ultralong,
	title={Ultralong spin-coherence times for rubidium atoms in solid parahydrogen via dynamical decoupling},
	author={Upadhyay, Sunil and Dargyte, Ugne and Patterson, David and Weinstein, Jonathan D},
	journal={Physical Review Letters},
	volume={125},
	number={4},
	pages={043601},
	year={2020},
	doi = "10.1103/PhysRevLett.125.043601",
	publisher={APS}
}

@article{lukin2001controlling,
	title={Controlling photons using electromagnetically induced transparency},
	author={Lukin, MD and Imamo{\u{g}}lu, Ata{\c{c}}},
	journal={Nature},
	volume={413},
	number={6853},
	pages={273--276},
	year={2001},
	publisher={Nature Publishing Group UK London}
}

@article{caldwell2020long,
	title={Long rotational coherence times of molecules in a magnetic trap},
	author={Caldwell, L and Williams, HJ and Fitch, NJ and Aldegunde, J and Hutson, Jeremy M and Sauer, BE and Tarbutt, MR},
	journal={Physical Review Letters},
	volume={124},
	number={6},
	pages={063001},
	year={2020},
	publisher={APS}
}

@article{marangos1998electromagnetically,
	title={Electromagnetically induced transparency},
	author={Marangos, Jonathan P},
	journal={Journal of modern optics},
	volume={45},
	number={3},
	pages={471--503},
	year={1998},
	publisher={Taylor \& Francis}
}

@article{harris1997electromagnetically,
	title={Electromagnetically induced transparency},
	author={Harris, Stephen E},
	journal={Physics today},
	volume={50},
	number={7},
	pages={36--42},
	year={1997},
	publisher={American Institute of Physics}
}

@article{kosachiov2000efficient,
	title={Efficient microwave-induced optical frequency conversion},
	author={Kosachiov, DV and Korsunsky, EA},
	journal={The European Physical Journal D-Atomic, Molecular, Optical and Plasma Physics},
	volume={11},
	number={3},
	pages={457--463},
	year={2000},
	publisher={Springer}
}

@article{basler2015radio,
	title={Radio-frequency-assisted electromagnetically induced transparency},
	author={Basler, Carl and Grzesiak, Jonas and Helm, Hanspeter},
	journal={Physical Review A},
	volume={92},
	number={1},
	pages={013809},
	year={2015},
	publisher={APS}
}

@article{vogt2018microwave,
	title={Microwave-assisted Rydberg electromagnetically induced transparency},
	author={Vogt, Thibault and Gross, Christian and Gallagher, Thomas F and Li, Wenhui},
	journal={Optics letters},
	volume={43},
	number={8},
	pages={1822--1825},
	year={2018},
	publisher={Optical Society of America}
}

@inproceedings{li2005control,
	title={Control an electromagnetically induced transparence feature with a microwave field},
	author={Li, Xiaoli and Zhang, Lianshui and Yang, Lijun and Feng, Xiaomin and Han, Li and Fu, Guangsheng and Manson, Neil B and Wei, Changjiang},
	booktitle={Quantum Optics and Applications in Computing and Communications II},
	volume={5631},
	pages={205--213},
	year={2005},
	organization={SPIE} 
}

@article{li2009electromagnetically,  
	title={Electromagnetically induced transparency controlled by a microwave field},
	author={Li, Hebin and Sautenkov, Vladimir A and Rostovtsev, Yuri V and Welch, George R and Hemmer, Philip R and Scully, Marlan O},
	journal={Physical Review A—Atomic, Molecular, and Optical Physics},
	volume={80},
	number={2},
	pages={023820},
	year={2009},
	publisher={APS}
}

@article{kuznetsova2002atomic,
	title={Atomic interference phenomena in solids with a long-lived spin coherence},
	author={Kuznetsova, Elena and Kocharovskaya, Olga and Hemmer, Philip and Scully, Marlan O},
	journal={Physical Review A},
	volume={66},
	number={6},
	pages={063802},
	year={2002},
	doi = "10.1103/PhysRevA.66.063802",
	publisher={APS}
}

@article{fleischhauer2005EIT,
	title={Electromagnetically induced transparency: Optics in coherent media},
	author={Fleischhauer, Michael and Imamoglu, Atac and Marangos, Jonathan P},
	journal={Reviews of modern physics},
	volume={77},
	number={2},
	pages={633--673},
	year={2005},
	doi = "10.1103/RevModPhys.77.633",
	publisher={APS}
}

@article{DUNE:2020lwj, author = "Abi, Babak and others", collaboration = "DUNE", title = "{Deep Underground Neutrino Experiment (DUNE), Far Detector Technical Design Report, Volume I Introduction to DUNE}", eprint = "2002.02967", archivePrefix = "arXiv", primaryClass = "physics.ins-det", reportNumber = "FERMILAB-PUB-20-024-ND, FERMILAB-DESIGN-2020-01", doi = "10.1088/1748-0221/15/08/T08008", journal = "JINST", volume = "15", number = "08", pages = "T08008", year = "2020" }

@article{DUNE:2020ypp, author = "Abi, Babak and others", collaboration = "DUNE", title = "{Deep Underground Neutrino Experiment (DUNE), Far Detector Technical Design Report, Volume II: DUNE Physics}", eprint = "2002.03005", archivePrefix = "arXiv", primaryClass = "hep-ex", reportNumber = "FERMILAB-PUB-20-025-ND, FERMILAB-DESIGN-2020-02", month = "2", year = "2020" }

@article{Essig:2019kfe,
	author = "Essig, Rouven and P{\'e}rez-R{\'\i}os, Jes{\'u}s and Ramani, Harikrishnan and Slone, Oren",
	title = "{Direct Detection of Spin-(In)dependent Nuclear Scattering of Sub-GeV Dark Matter Using Molecular Excitations}",
	eprint = "1907.07682",
	archivePrefix = "arXiv",
	primaryClass = "hep-ph",
	doi = "10.1103/PhysRevResearch.1.033105",
	journal = "Phys. Rev. Research.",
	volume = "1",
	pages = "033105",
	year = "2019"
}

@article{JNE:2020bwn,
	author = "Guo, Ziyi and others",
	collaboration = "JNE",
	title = "{Muon flux measurement at China Jinping Underground Laboratory}",
	eprint = "2007.15925",
	archivePrefix = "arXiv",
	primaryClass = "physics.ins-det",
	doi = "10.1088/1674-1137/abccae",
	journal = "Chin. Phys. C",
	volume = "45",
	number = "2",
	pages = "025001",
	year = "2021"
}

@article{Akhmedov:2018wlf,
	author = "Akhmedov, Evgeny and Arcadi, Giorgio and Lindner, Manfred and Vogl, Stefan",
	title = "{Coherent scattering and macroscopic coherence: Implications for neutrino, dark matter and axion detection}",
	eprint = "1806.10962",
	archivePrefix = "arXiv",
	primaryClass = "hep-ph",
	doi = "10.1007/JHEP10(2018)045",
	journal = "JHEP",
	volume = "10",
	pages = "045",
	year = "2018"
}

@ARTICLE{2018NatAs,
	author = {{Echternach}, P.~M. and {Pepper}, B.~J. and {Reck}, T. and {Bradford}, C.~M.},
	title = "{Single photon detection of 1.5 THz radiation with the quantum capacitance detector}",
	journal = {Nature Astronomy},
	year = 2018,
	volume = {2},
	pages = {90-97},
	doi = {10.1038/s41550-017-0294-y}
}

@article{Homma:2024aan,
	author = "Homma, Kensuke and Kirita, Yuri and Miyamaru, Takafumi and Hasada, Takumi and Kodama, Airi",
	title = "{Opening a meV mass window for axionlike particles with a microwave-laser-mixed stimulated resonant photon collider}",
	eprint = "2405.03577",
	archivePrefix = "arXiv",
	primaryClass = "hep-ph",
	doi = "10.1103/PhysRevD.110.092017",
	journal = "Phys. Rev. D",
	volume = "110",
	number = "9",
	pages = "092017",
	year = "2024"
}

@article{Kim:2018jtj,
	author = "Kim, S. H. and others",
	title = "{Development of Superconducting Tunnel Junction Far-Infrared Photon Detector for Cosmic Background Neutrino Decay Search - COBAND experiment}",
	reportNumber = "FERMILAB-CONF-18-836-PPD",
	doi = "10.22323/1.340.0427",
	journal = "PoS",
	volume = "ICHEP2018",
	pages = "427",
	year = "2019"
}

@article{Baselmans:2016pdc,
	author = "Baselmans, J. J. A. and others",
	title = "{A kilo-pixel imaging system for future space based far-infrared observatories using microwave kinetic inductance detectors}",
	eprint = "1609.01952",
	archivePrefix = "arXiv",
	primaryClass = "astro-ph.IM",
	doi = "10.1051/0004-6361/201629653",
	journal = "Astron. Astrophys.",
	volume = "601",
	pages = "A89",
	year = "2017"
}

@article{Bouchiat:1975zz,
	author = "Bouchiat, M. A. and Bouchiat, C.",
	title = "{Parity Violation Induced by Weak Neutral Currents in Atomic Physics. Part 2}",
	doi = "10.1051/jphys:01975003606049300",
	journal = "J. Phys. (France)",
	volume = "36",
	pages = "493--509",
	year = "1975"
}

@article{BOUCHIAT1982358,
	title = {Observation of a parity violation in cesium},
	journal = {Physics Letters B},
	volume = {117},
	number = {5},
	pages = {358-364},
	year = {1982},
	issn = {0370-2693},
	doi = {https://doi.org/10.1016/0370-2693(82)90736-5},
	url = {https://www.sciencedirect.com/science/article/pii/0370269382907365},
	author = {M.A. Bouchiat and J. Guena and L. Hunter and L. Pottier}
}

@book{Bransden2006,
	title =     {Physics of Atoms and Molecules},
	author =    {B.H. Bransden and C.J. Joachain},
	publisher = {Dorling Kindersley Pvt Ltd},
	year =      {2006},
}

@article{Safronova:2017xyt,
	author = "Safronova, M. S. and Budker, D. and DeMille, D. and Kimball, Derek F. Jackson and Derevianko, A. and Clark, C. W.",
	title = "{Search for New Physics with Atoms and Molecules}",
	eprint = "1710.01833",
	archivePrefix = "arXiv",
	primaryClass = "physics.atom-ph",
	doi = "10.1103/RevModPhys.90.025008",
	journal = "Rev. Mod. Phys.",
	volume = "90",
	number = "2",
	pages = "025008",
	year = "2018"
}

@article{Bauer:2024dbg,
	author = "Bauer, Martin and Perez-Soler, Javier and Shergold, Jack D.",
	title = "{Generalised hydrogen interactions with CINCO: a window to new physics}",
	eprint = "2407.12913",
	archivePrefix = "arXiv",
	primaryClass = "hep-ph",
	doi = "10.1007/JHEP10(2024)176",
	journal = "JHEP",
	volume = "10",
	pages = "176",
	year = "2024"
}

@article{Bhoonah:2019eyo,
	author = "Bhoonah, Amit and Bramante, Joseph and Song, Ningqiang",
	title = "{Superradiant Searches for Dark Photons in Two Stage Atomic Transitions}",
	eprint = "1909.07387",
	archivePrefix = "arXiv",
	primaryClass = "hep-ph",
	doi = "10.1103/PhysRevD.101.055040",
	journal = "Phys. Rev. D",
	volume = "101",
	number = "5",
	pages = "055040",
	year = "2020"
}

@article{Ge:2021lur,
	author = "Ge, Shao-Feng and Pasquini, Pedro",
	title = "{Probing light mediators in the radiative emission of neutrino pair}",
	eprint = "2110.03510",
	archivePrefix = "arXiv",
	primaryClass = "hep-ph",
	doi = "10.1140/epjc/s10052-022-10131-4",
	journal = "Eur. Phys. J. C",
	volume = "82",
	number = "3",
	pages = "208",
	year = "2022"
}

@article{Ge:2022cib,
	author = "Ge, Shao-Feng and Pasquini, Pedro",
	title = "{Unique probe of neutrino electromagnetic moments with radiative pair emission}",
	eprint = "2206.11717",
	archivePrefix = "arXiv",
	primaryClass = "hep-ph",
	doi = "10.1016/j.physletb.2023.137911",
	journal = "Phys. Lett. B",
	volume = "841",
	pages = "137911",
	year = "2023"
}

@article{Ge:2023oag,
	author = "Ge, Shao-Feng and Pasquini, Pedro",
	title = "{Disentangle neutrino electromagnetic properties with atomic radiative pair emission}",
	eprint = "2306.12953",
	archivePrefix = "arXiv",
	primaryClass = "hep-ph",
	doi = "10.1007/JHEP12(2023)083",
	journal = "JHEP",
	volume = "12",
	pages = "083",
	year = "2023"
}

@article{Huang:2019phr,
	author = "Huang, Guo-Yuan and Sasao, Noboru and Xing, Zhi-Zhong and Yoshimura, Motohiko",
	title = "{Testing unitarity of the $3\times 3$ neutrino mixing matrix in an atomic system}",
	eprint = "1904.10366",
	archivePrefix = "arXiv",
	primaryClass = "hep-ph",
	doi = "10.1142/S0217751X20500049",
	journal = "Int. J. Mod. Phys. A",
	volume = "35",
	number = "01",
	pages = "2050004",
	year = "2020"
}

@article{Zhang:2016lqp,
	author = "Zhang, Jue and Zhou, Shun",
	title = "{Improved Statistical Determination of Absolute Neutrino Masses via Radiative Emission of Neutrino Pairs from Atoms}",
	eprint = "1604.08008",
	archivePrefix = "arXiv",
	primaryClass = "hep-ph",
	doi = "10.1103/PhysRevD.93.113020",
	journal = "Phys. Rev. D",
	volume = "93",
	number = "11",
	pages = "113020",
	year = "2016"
}

@article{Huang:2019rmc,
	author = "Huang, Guo-Yuan and Zhou, Shun",
	title = "{Probing Cosmic Axions through Resonant Emission and Absorption in Atomic Systems with Superradiance}",
	eprint = "1905.00367",
	archivePrefix = "arXiv",
	primaryClass = "hep-ph",
	doi = "10.1103/PhysRevD.100.035010",
	journal = "Phys. Rev. D",
	volume = "100",
	number = "3",
	pages = "035010",
	year = "2019"
}

@article{Arvanitaki:2024taq,
	author = "Arvanitaki, Asimina and Dimopoulos, Savas and Galanis, Marios",
	title = "{Superradiant interactions of the cosmic neutrino background, axions, dark matter, and reactor neutrinos}",
	eprint = "2408.04021",
	archivePrefix = "arXiv",
	primaryClass = "hep-ph",
	doi = "10.1103/PhysRevD.111.055015",
	journal = "Phys. Rev. D",
	volume = "111",
	number = "5",
	pages = "055015",
	year = "2025"
}

@article{Arvanitaki:2017nhi,
	author = "Arvanitaki, Asimina and Dimopoulos, Savas and Van Tilburg, Ken",
	title = "{Resonant absorption of bosonic dark matter in molecules}",
	eprint = "1709.05354",
	archivePrefix = "arXiv",
	primaryClass = "hep-ph",
	doi = "10.1103/PhysRevX.8.041001",
	journal = "Phys. Rev. X",
	volume = "8",
	number = "4",
	pages = "041001",
	year = "2018"
}

@article{Yoshimura:2006nd,
	author = "Yoshimura, M.",
	title = "{Neutrino Pair Emission from Excited Atoms}",
	eprint = "hep-ph/0611362",
	archivePrefix = "arXiv",
	doi = "10.1103/PhysRevD.75.113007",
	journal = "Phys. Rev. D",
	volume = "75",
	pages = "113007",
	year = "2007"
}

@article{Yoshimura:2012tm,
	author = "Yoshimura, M. and Sasao, N. and Tanaka, M.",
	title = "{Dynamics of paired superradiance}",
	eprint = "1203.5394",
	archivePrefix = "arXiv",
	primaryClass = "quant-ph",
	reportNumber = "OU-HET-763-2012, OU-HET 763/2012",
	doi = "10.1103/PhysRevA.86.013812",
	journal = "Phys. Rev. A",
	volume = "86",
	pages = "013812",
	year = "2012"
}

@article{Dinh:2012qb,
	author = "Dinh, D. N. and Petcov, S. T. and Sasao, N. and Tanaka, M. and Yoshimura, M.",
	title = "{Observables in Neutrino Mass Spectroscopy Using Atoms}",
	eprint = "1209.4808",
	archivePrefix = "arXiv",
	primaryClass = "hep-ph",
	doi = "10.1016/j.physletb.2013.01.015",
	journal = "Phys. Lett. B",
	volume = "719",
	pages = "154--163",
	year = "2013"
}

@article{Song:2015xaa,
	author = "Song, Ningqiang and Boyero Garcia, R. and Gomez-Cadenas, J. J. and Gonzalez-Garcia, M. C. and Peralta Conde, A. and Taron, Josep",
	title = "{Conditions for Statistical Determination of the Neutrino Mass Spectrum in Radiative Emission of Neutrino Pairs in Atoms}",
	eprint = "1510.00421",
	archivePrefix = "arXiv",
	primaryClass = "hep-ph",
	reportNumber = "YITP-SB-15-38",
	doi = "10.1103/PhysRevD.93.013020",
	journal = "Phys. Rev. D",
	volume = "93",
	number = "1",
	pages = "013020",
	year = "2016"
}

@article{Fukumi:2012rn,
	author = "Fukumi, Atsushi and others",
	title = "{Neutrino Spectroscopy with Atoms and Molecules}",
	eprint = "1211.4904",
	archivePrefix = "arXiv",
	primaryClass = "hep-ph",
	reportNumber = "OU-HET-757-2012",
	doi = "10.1093/ptep/pts066",
	journal = "PTEP",
	volume = "2012",
	pages = "04D002",
	year = "2012"
}

@article{Yoshimura:2014hfa,
	author = "Yoshimura, M. and Sasao, N. and Tanaka, M.",
	title = "{Experimental method of detecting relic neutrino by atomic de-excitation}",
	eprint = "1409.3648",
	archivePrefix = "arXiv",
	primaryClass = "hep-ph",
	reportNumber = "OU-HET-826",
	doi = "10.1103/PhysRevD.91.063516",
	journal = "Phys. Rev. D",
	volume = "91",
	number = "6",
	pages = "063516",
	year = "2015"
}

@book{Khurgin2009,
	title =     {Slow Light: Science and Applications},
	author =    {Jacob B. Khurgin, Rodney S. Tucker},
	publisher = {CRC Press},
	isbn =      {1420061518; 9781420061512; 9781420061529; 1420061526},
	year =      {2008},
	series =    {Optical Science and Engineering},
	edition =   {1},
}

@article{Khurgin2010,
	author = {Jacob B. Khurgin},
	journal = {Adv. Opt. Photon.},
	number = {3},
	pages = {287--318},
	publisher = {Optica Publishing Group},
	title = {Slow light in various media: a tutorial},
	volume = {2},
	year = {2010},
	url = {https://opg.optica.org/aop/abstract.cfm?URI=aop-2-3-287},
	doi = {10.1364/AOP.2.000287}
}

@article{hau1999light,
	title={Light speed reduction to 17 metres per second in an ultracold atomic gas},
	author={Hau, Lene Vestergaard and Harris, Stephen E and Dutton, Zachary and Behroozi, Cyrus H},
	journal={Nature},
	volume={397},
	number={6720},
	pages={594--598},
	year={1999},
	publisher={Nature Publishing Group UK London}
}

@article{JUNO:2015zny,
	author = "An, Fengpeng and others",
	collaboration = "JUNO",
	title = "{Neutrino Physics with JUNO}",
	eprint = "1507.05613",
	archivePrefix = "arXiv",
	primaryClass = "physics.ins-det",
	doi = "10.1088/0954-3899/43/3/030401",
	journal = "J. Phys. G",
	volume = "43",
	number = "3",
	pages = "030401",
	year = "2016"
}

@article{JUNO:2024jaw,
	author = "Abusleme, Angel and others",
	collaboration = "JUNO",
	title = "{Potential to identify neutrino mass ordering with reactor antineutrinos at JUNO}",
	eprint = "2405.18008",
	archivePrefix = "arXiv",
	primaryClass = "hep-ex",
	doi = "10.1088/1674-1137/ad7f3e",
	journal = "Chin. Phys. C",
	volume = "49",
	number = "3",
	pages = "033104",
	year = "2025"
}

@article{Ruzi:2023cvp,
	author = "Ruzi, Alim and Qian, Sitian and Yang, Tianyi and Li, Qiang",
	title = "{Low Energy Neutrino and Mass Dark Matter Detection Using Freely Falling Atoms}",
	eprint = "2302.09874",
	archivePrefix = "arXiv",
	primaryClass = "hep-ph",
	month = "2",
	year = "2023"
}

@article{Planck:2018vyg,
	author = "Aghanim, N. and others",
	collaboration = "Planck",
	title = "{Planck 2018 results. VI. Cosmological parameters}",
	eprint = "1807.06209",
	archivePrefix = "arXiv",
	primaryClass = "astro-ph.CO",
	doi = "10.1051/0004-6361/201833910",
	journal = "Astron. Astrophys.",
	volume = "641",
	pages = "A6",
	year = "2020",
	note = "[Erratum: Astron.Astrophys. 652, C4 (2021)]"
}

@book{Boyd2008,
	author = {Boyd, Robert W.},
	title = {Nonlinear Optics, Third Edition},
	year = {2008},
	isbn = {0123694701},
	publisher = {Academic Press, Inc.},
	address = {USA},
	edition = {3rd}
}

@book{Butcher_Cotter_1990, place={Cambridge}, series={Cambridge Studies in Modern Optics}, title={The Elements of Nonlinear Optics}, publisher={Cambridge University Press}, author={Butcher, Paul N. and Cotter, David}, year={1990}, collection={Cambridge Studies in Modern Optics}}

@book{Klyshko1988,
	title =     {Photons and nonlinear optics},
	author =    {D  N Klyshko},
	publisher = {Gordon and Breach },
	isbn =      {2881246699,9782881246692},
	year =      {1988},
	series =    {},
	edition =   {Rev. and enl. ed},
	volume =    {},
	url =       {http://gen.lib.rus.ec/book/index.php?md5=e2023ecda1989654549c96589a22a8a2}
}

@article{YanhuaShih2003,
	doi = {10.1088/0034-4885/66/6/203},
	url = {https://dx.doi.org/10.1088/0034-4885/66/6/203},
	year = {2003},
	volume = {66},
	number = {6},
	pages = {1009},
	author = {Yanhua Shih},
	title = {Entangled biphoton source - property and preparation},
	journal = {Reports on Progress in Physics}
}

@article{PTOLEMY:2018jst,
	author = "Baracchini, E. and others",
	collaboration = "PTOLEMY",
	title = "{PTOLEMY: A Proposal for Thermal Relic Detection of Massive Neutrinos and Directional Detection of MeV Dark Matter}",
	eprint = "1808.01892",
	archivePrefix = "arXiv",
	primaryClass = "physics.ins-det",
	month = "8",
	year = "2018"
}

@article{PTOLEMY:2019hkd,
	author = "Betti, M. G. and others",
	collaboration = "PTOLEMY",
	title = "{Neutrino physics with the PTOLEMY project: active neutrino properties and the light sterile case}",
	eprint = "1902.05508",
	archivePrefix = "arXiv",
	primaryClass = "astro-ph.CO",
	doi = "10.1088/1475-7516/2019/07/047",
	journal = "JCAP",
	volume = "07",
	pages = "047",
	year = "2019"
}

@article{PTOLEMY:2022ldz,
	author = "Apponi, A. and others",
	collaboration = "PTOLEMY",
	title = "{Heisenberg\textquoteright{}s uncertainty principle in the PTOLEMY project: A theory update}",
	eprint = "2203.11228",
	archivePrefix = "arXiv",
	primaryClass = "hep-ph",
	doi = "10.1103/PhysRevD.106.053002",
	journal = "Phys. Rev. D",
	volume = "106",
	number = "5",
	pages = "053002",
	year = "2022"
}

@article{Cheipesh:2023qiy,
	author = "Cheipesh, Yevheniia and Ridkokasha, Ivan and Cheianov, Vadim and Boyarsky, Alexey",
	title = "{Can we really detect relic neutrinos?}",
	doi = "10.21468/SciPostPhysProc.12.042",
	journal = "SciPost Phys. Proc.",
	volume = "12",
	pages = "042",
	year = "2023"
}

@article{Cheipesh:2021fmg,
	author = "Cheipesh, Yevheniia and Cheianov, Vadim and Boyarsky, Alexey",
	title = "{Navigating the pitfalls of relic neutrino detection}",
	eprint = "2101.10069",
	archivePrefix = "arXiv",
	primaryClass = "hep-ph",
	doi = "10.1103/PhysRevD.104.116004",
	journal = "Phys. Rev. D",
	volume = "104",
	number = "11",
	pages = "116004",
	year = "2021"
}

@article{Weinberg:1962zza,
	author = "Weinberg, Steven",
	title = "{Universal Neutrino Degeneracy}",
	doi = "10.1103/PhysRev.128.1457",
	journal = "Phys. Rev.",
	volume = "128",
	pages = "1457--1473",
	year = "1962"
}

@article{Shergold:2021evs,
    author = "Shergold, Jack D.",
    title = "{Updated detection prospects for relic neutrinos using coherent scattering}",
    eprint = "2109.07482",
    archivePrefix = "arXiv",
    primaryClass = "hep-ph",
    reportNumber = "IPPP/21/28",
    doi = "10.1088/1475-7516/2021/11/052",
    journal = "JCAP",
    volume = "11",
    number = "11",
    pages = "052",
    year = "2021"
}

@article{Opher:1974drq,
	author = "Opher, R.",
	title = "{Coherent scattering of cosmic neutrinos}",
	journal = "Astron. Astrophys.",
	volume = "37",
	number = "1",
	pages = "135--137",
	year = "1974"
}

@article{Ringwald:2009bg,
	author = "Ringwald, Andreas",
	editor = "Tserruya, Itzhak and Gal, Avraham and Ashery, Daniel",
	title = "{Prospects for the direct detection of the cosmic neutrino background}",
	eprint = "0901.1529",
	archivePrefix = "arXiv",
	primaryClass = "astro-ph.CO",
	reportNumber = "DESY-09-001",
	doi = "10.1016/j.nuclphysa.2009.05.109",
	journal = "Nucl. Phys. A",
	volume = "827",
	pages = "501C--506C",
	year = "2009"
}

@article{Gelmini:2004hg,
	author = "Gelmini, Graciela B.",
	editor = {Bergstr\"om, L. and Botner, O. and Carlson, P. and Hulth, P. O. and Ohlsson, T.},
	title = "{Prospect for relic neutrino searches}",
	eprint = "hep-ph/0412305",
	archivePrefix = "arXiv",
	reportNumber = "UCLA-04-TEP-53",
	doi = "10.1088/0031-8949/2005/T121/019",
	journal = "Phys. Scripta T",
	volume = "121",
	pages = "131--136",
	year = "2005"
}

@article{Duda:2001hd,
	author = "Duda, Gintaras and Gelmini, Graciela and Nussinov, Shmuel",
	title = "{Expected signals in relic neutrino detectors}",
	eprint = "hep-ph/0107027",
	archivePrefix = "arXiv",
	reportNumber = "UCLA-01-TEP-6",
	doi = "10.1103/PhysRevD.64.122001",
	journal = "Phys. Rev. D",
	volume = "64",
	pages = "122001",
	year = "2001"
}

@article{Cabibbo:1982bb,
	author = "Cabibbo, N. and Maiani, L.",
	title = "{The Vanishing of Order $G$ Mechanical Effects of Cosmic Massive Neutrinos on Bulk Matter}",
	reportNumber = "Print-82-0419 (ROME)",
	doi = "10.1016/0370-2693(82)90127-7",
	journal = "Phys. Lett. B",
	volume = "114",
	pages = "115--117",
	year = "1982"
}

@article{Lewis:1979mu,
	author = "Lewis, R. R.",
	title = "{Coherent Detector for Low-energy Neutrinos}",
	reportNumber = "ITP-SB-79-39",
	doi = "10.1103/PhysRevD.21.663",
	journal = "Phys. Rev. D",
	volume = "21",
	pages = "663",
	year = "1980"
}

@article{Smith:1983jj,
	author = "Smith, P. F. and Lewin, J. D.",
	title = "{Coherent Interaction of Galactic Neutrinos with Material Targets}",
	doi = "10.1016/0370-2693(83)90873-0",
	journal = "Phys. Lett. B",
	volume = "127",
	pages = "185--190",
	year = "1983"
}

@article{Shvartsman:1982sn,
	author = "Shvartsman, B. F. and Braginsky, V. B. and Gershtein, S. S. and Zeldovich, Ya. B. and Khlopov, M. Yu.",
	title = "{Possibility of Detecting Relic Massive Neutrinos}",
	journal = "JETP Lett.",
	volume = "36",
	pages = "277--279",
	year = "1982"
}

@article{Domcke:2017aqj,
	author = "Domcke, Valerie and Spinrath, Martin",
	title = "{Detection prospects for the Cosmic Neutrino Background using laser interferometers}",
	eprint = "1703.08629",
	archivePrefix = "arXiv",
	primaryClass = "astro-ph.CO",
	reportNumber = "NCTS-PH-1702",
	doi = "10.1088/1475-7516/2017/06/055",
	journal = "JCAP",
	volume = "06",
	pages = "055",
	year = "2017"
}

@article{Zeldovich:1981wf,
	author = "Zeldovich, Ya. B. and Khlopov, M. Yu.",
	title = "{The Neutrino Mass in Elementary Particle Physics and in Big Bang Cosmology}",
	doi = "10.1070/PU1981v024n09ABEH004816",
	journal = "Sov. Phys. Usp.",
	volume = "24",
	pages = "755--774",
	year = "1981"
}

@article{Langacker:1982ih,
	author = "Langacker, Paul and Leveille, Jacques P. and Sheiman, Jon",
	title = "{On the Detection of Cosmological Neutrinos by Coherent Scattering}",
	reportNumber = "UM HE 82-28",
	doi = "10.1103/PhysRevD.27.1228",
	journal = "Phys. Rev. D",
	volume = "27",
	pages = "1228",
	year = "1983"
}

@article{Lewis:1987yd,
	author = "Lewis, R. R.",
	title = "{Radiation Pressure of Neutrinos in Refracting Media}",
	doi = "10.1103/PhysRevD.35.2134",
	journal = "Phys. Rev. D",
	volume = "35",
	pages = "2134--2141",
	year = "1987"
}

@article{Loeb:1990xs,
	author = "Loeb, Abraham and Starkman, Glenn D.",
	title = "{A Detector for the Cosmic Neutrino Background}",
	reportNumber = "IASSNS-AST-90-10",
	doi = "10.1016/0920-5632(91)90205-S",
	journal = "Nucl. Phys. B Proc. Suppl.",
	volume = "19",
	pages = "241--250",
	year = "1991"
}

@article{Ferreras:1995wf,
	author = "Ferreras, I. and Wasserman, I.",
	title = "{Feasibility of observing mechanical effects of cosmological neutrinos}",
	doi = "10.1103/PhysRevD.52.5459",
	journal = "Phys. Rev. D",
	volume = "52",
	pages = "5459--5479",
	year = "1995"
}

@article{Vogel:2015vfa,
	author = "Vogel, Petr",
	editor = "Kearns, Ed",
	title = "{How difficult it would be to detect cosmic neutrino background?}",
	doi = "10.1063/1.4915587",
	journal = "AIP Conf. Proc.",
	volume = "1666",
	number = "1",
	pages = "140003",
	year = "2015"
}

@inproceedings{Hagmann:1999kf,
	author = "Hagmann, C.",
	title = "{Cosmic neutrinos and their detection}",
	booktitle = "{American Physical Society (APS) Meeting of the Division of Particles and Fields (DPF 99)}",
	eprint = "astro-ph/9905258",
	archivePrefix = "arXiv",
	reportNumber = "UCRL-JC-134163",
	month = "1",
	year = "1999"
}

\clearpage

\onecolumngrid

\section*{\large Supplemental Material of ``Probing Cosmic Neutrino Background through Parametric Fluorescence''}

\subsection{Pure Electromagnetic Processes} \label{appendix:A}

\begin{figure}[t!]
	\begin{center}
		\includegraphics[width=0.48\textwidth]{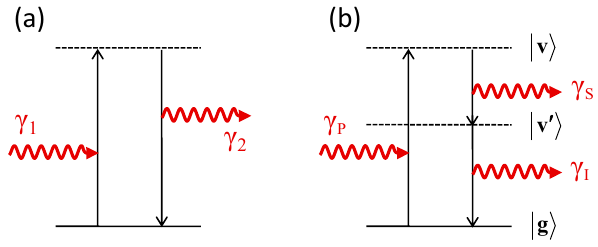}
	\end{center}
	\caption{The fundamental diagrams for the generation of additional electromagnetic couplings by coherently scattering with atomic/molecular energy levels: (a) the coupling $\gamma\gamma$ corresponding to the linear electric susceptibility which changes the refractive index of lights; {(b)} the coupling $\gamma\gamma\gamma$ corresponding to the nonlinear electric susceptibility that induces the down conversion of pumping photons. The ground and virtual states are denoted by $\left|\rm g \right>$ and $\left|\rm v \right>$ (or $\left|\rm v^\prime \right>$), respectively. Those scatterings are coherent because the transition  returns back to the ground state via the short-lived intermediate state.
	}
	\label{fig:scheme2}
\end{figure}

It will be instructive to review the pure electromagnetic processes in nonlinear optics~\cite{Boyd2008} and outline the necessary procedure in accordance with the notations of the present work.
For a dielectric medium, the polarization can be formally expanded as power series of the electric field in the weak-field approximation,
\begin{align}
	{P}^{}_{i} \supset \chi^{(1)}_{i j} {\mathcal{E}}^{}_{j} + \chi^{(2)}_{i j k} {\mathcal{E}}^{}_{j} {\mathcal{E}}^{}_{k} + \chi^{(3)}_{i j k l} {\mathcal{E}}^{}_{j} {\mathcal{E}}^{}_{k} {\mathcal{E}}^{}_{l} + \cdots \;,
\end{align}
where $\chi^{(1)}$ is known as the linear susceptibility of the medium, $\chi^{(i)}$ (for $i > 1$)  denotes the nonlinear susceptibilities, {and $\mathcal{E}^{}_{j}$ (for $j = 1,2,3$) stands for the electric field.}
The linear susceptibility, corresponding to the diagram in Fig.~(\ref{fig:scheme2}a), is connected to the refractive index through $n = \mathrm{Re} (\sqrt{1+\chi^{(1)}})$.
The SPDC process is controlled by $\chi^{(2)}$ describing three photon couplings, which is a second-order process. 
Depicted by Fig.~(\ref{fig:scheme2}b), SPDC requires a three-level system, for which
different levels are connected via the electric-dipole interaction $H^{}_{\bm{d}} = -\bm{d} \cdot \bm{\mathcal{E}}$.

Suppose that the electromagnetic field is composed of plane waves with several discrete frequencies, $\bm{\mathcal{E}}(t, \bm{x}) = \sum^{}_{i}\bm{\mathcal{E}}(\bm{k}^{}_{i}) \cdot \mathrm{exp}(-i \omega^{}_{i} t )$ including both the positive and negative frequencies. Note that the spatial phase is implicitly incorporated in $\bm{\mathcal{E}}(\bm{k}^{}_{i})$.
Under the perturbation of this field, the level coefficient of the virtual state $\left|  \rm v \right>$ to the first order evolve as 
\begin{align}
	C^{(1)}_{\rm v} (t)  & = -\mathrm{i} \int^{t}_{-\infty} \mathrm{d}\tilde{t} \left<{\rm v} \right| H^{}_{\bm{d}}(\tilde{t}) \left| {\rm g} \right> = \sum^{}_{i} \frac{\bm{d}^{}_{\rm v g} \cdot \bm{\mathcal{E}}(\omega^{}_{i}) }{E^{}_{\rm vg} - \omega^{}_{i}} \mathrm{e}^{\mathrm{i} (E^{}_{\rm vg} - \omega^{}_{i}) t}\;,
\end{align}
where we have neglected the rapidly oscillating term when the time integration approaches infinity. The linear polarization per radiant $\left< \bm{d}\right> =  \sum^{}_{\rm v} C^{(1)}_{\rm v} \left<\rm g \right|\bm{d} \left|\rm  v\right> + C^{(1) *}_{\rm v} \left<\rm v \right|\bm{d} \left| \rm g\right> $ summing over all the virtual states has the following explicit expression
\begin{align} \label{eq:d1}
	\left< \bm{d}\right>^{(1)}  &  = \sum^{}_{i,\, {\rm v}} \frac{ \bm{d}^{*}_{\rm vg} \left[ \bm{d}^{}_{\rm vg} \cdot \bm{\mathcal{E}}(\omega^{}_{i}) \right] }{E^{}_{\rm vg} - \omega^{}_{i}} \mathrm{e}^{ - \mathrm{i}  \omega^{}_{i} t} + {\rm h.c.} 
\end{align}
The polarization per unit volume then reads $\bm{P}^{(1)} = n^{}_{\bm{d}}\,\left< \bm{d}\right>^{(1)}$, where $n^{}_{\bm{d}} \sim N^{}_{\rm A}~{\rm cm^{-3}}$ represents the density of dipoles. 
Diagrammatically, this first-order polarization stems from the transition process $\left|\rm g \right> \overset{\gamma}{\to} \left|\rm v \right> \overset{\gamma}{\to} \left|\rm g \right>$ as in Fig.~(\ref{fig:scheme2}b).
The first term (or its hermitian conjugate) of Eq.~(\ref{eq:d1}) corresponds to the scenario where $\gamma^{}_{1}$ (or $\gamma^{}_{2}$) is taken as the primary perturbation.
Using $\mathcal{H}^{(1)}_{\rm 2\gamma} = - \bm{P}^{(1)} \cdot \bm{\mathcal{E}}(t,\bm{x})$, the Hamiltonian density due to the medium effect reads
\begin{align} \label{eq:}
	\mathcal{H}^{(1)}_{\rm 2\gamma}  = & \sum^{}_{i,\,j,\, {\rm v}} \frac{n^{}_{\bm{d}}\left[\bm{d}^{*}_{\rm vg} \cdot \bm{\mathcal{E}}(\bm{k}^{}_{j}) \right] \left[ \bm{d}^{}_{\rm v g} \cdot \bm{\mathcal{E}}(\bm{k}^{}_{i}) \right] }{\omega^{}_{i} -E^{}_{\rm vg}} \mathrm{e}^{ - \mathrm{i} ( \omega^{}_{i} + \omega^{}_{j}) t} + {\rm h.c.} \;,
\end{align}
which couples the electric components of two photons.

Let us now proceed to the second-order process.
The second-order coefficients under the perturbation can be obtained by a further iteration as 
\begin{align}
	C^{(2)}_{\rm v} (t)  & = -\mathrm{i} \int^{t}_{-\infty} \mathrm{d}\tilde{t} \sum^{}_{\rm v^\prime} \left<{\rm v} \right| H^{}_{\bm{d}}(\tilde{t}) \left| {\rm v^{\prime}} \right> C^{(1)}_{\rm v^\prime}(\tilde{t}) = \sum^{}_{i,\,j,\, {\rm v^\prime}} \frac{\left[ \bm{d}^{}_{\rm v v^\prime} \cdot \bm{\mathcal{E}}(\omega^{}_{j}) \right]\left[\bm{d}^{}_{\rm v^\prime g} \cdot \bm{\mathcal{E}}(\omega^{}_{i}) \right]}{(E^{}_{\rm vg} - \omega^{}_{i} - \omega^{}_{j}) (E^{}_{\rm v^\prime g} - \omega^{}_{i})} \mathrm{e}^{\mathrm{i} (E^{}_{\rm vg} - \omega^{}_{i} - \omega^{}_{j}) t}. 
\end{align}
Similar to the first-order case, the corresponding polarization per volume has the following expression 
\begin{align}
	\bm{P}^{(2)} & = n^{}_{\bm{d}} \left( \sum^{}_{\rm v} C^{(2)}_{\rm v} \left<\rm g \right|\bm{d} \left|\rm  v\right> + C^{(2)*}_{\rm v} \left<\rm v \right|\bm{d} \left|\rm  g\right>   + \sum^{}_{\rm v,\, v^\prime} C^{(1)}_{\rm v} C^{(1)*}_{\rm v^\prime} \left<\rm v^\prime \right|\bm{d} \left|\rm  v\right> \right) .
\end{align}
The second-order Hamiltonian density can be straightforwardly obtained with $\mathcal{H}^{(2)}_{\rm 3\gamma} = -\bm{P}^{(2)} \cdot \bm{\mathcal{E}}(t,\bm{x})$. 
However, the result of the Hamiltonian for the $\gamma\gamma\gamma$ coupling is quite tedious and involves  many permutations of photon fields, and thus we do not present the explicit expression here. 
With those results, one is able to calculate the rate for the SPDC process $\gamma^{}_{\rm P} \to \gamma^{}_{\rm S} + \gamma^{}_{\rm I}$, after the material has been specified.
The coherence requirement leads to a  phase matching condition, i.e., $\bm{k}^{}_{\rm P} = \bm{k}^{}_{\rm S} + \bm{k}^{}_{\rm I}$, along with the energy conservation. 
{For relic neutrino detection, there can similarly be second-order processes such as $\nu_i \to \nu_j + \gamma + \gamma$. Compared to the first-order case, these are  suppressed by an additional fine-structure constant. Nevertheless, it would be interesting to explore their potential effects in the future.}

\subsection{The Hamiltonian at the Atomic/Molecular Level}
Here, we derive the Hamiltonian describing the coupling between neutrinos and energy levels. To start, we first write down the four-fermion coupling for the non-diagonal transition of neutrinos,
\begin{eqnarray} \label{eq:Leff} 
	-\mathcal{L} & \supset &
	\sum^3_{i,j=1} \frac{G^{}_{\rm F} C^{}_{ij} }{\sqrt{2}} 
	\overline{\nu}^{}_{j} \gamma^{\mu} \left(1 -\gamma^{}_{5}\right)
	\nu^{}_{i} \cdot \overline{e}  \gamma^{}_{\mu}
	\left(1-\gamma^{}_{5}\right) e \;,\ \ \
\end{eqnarray}
where $G^{}_{\rm F}$ is the Fermi constant and $C^{}_{ij} \equiv U^*_{ei} U^{}_{ej}$ ($i\neq j$) with $U$ being the Pontecorvo-Maki-Nakagawa-Sakata (PMNS) matrix. The above Lagrangian can be simplified due to the following two facts: (i) electrons are non-relativistic in atoms; (ii) for an efficient electromagnetic transition from $\left| \rm v\right>$ to $\left| \rm g\right>$, the electron current should be chosen to be of either E1 or M1 type. Thus, we identify  two relevant parts, i.e., the temporal term of the vector current $\overline{e}  \gamma^{}_{0}
e$ and the spatial term of the axial-vector current $\overline{e}  \bm{\gamma}
\gamma^{}_{5} e$. 

The temporal term $\overline{e}  \gamma^{}_{0} e$ will lead to the E1-type transition. 
To clarify this, we denote the electron wavefunctions for the ground and virtual states as $\Psi^{}_{\rm g}$ and $\Psi^{}_{\rm v}$, respectively. The transition matrix $\left< \rm v \right| {H} \left| \rm g \right>$ will be proportional to 
$\int \mathrm{d}^3 \bm{x} \, \Psi^{*}_{\rm v}(\bm{x}) \Psi^{}_{\rm g} (\bm{x}) \mathrm{exp}[{\mathrm{i}(\bm{p}^{}_{i} - \bm{p}^{\prime}_{j}) \cdot \bm{x}}]$, where 
the spatial phase factor comes from the neutrino field decomposition. Expanding the phase factor, the first non-vanishing term is $\mathrm{i} \int \mathrm{d}^3 \bm{x} \, \Psi^{*}_{\rm v}(\bm{x}) \bm{x}  \Psi^{}_{\rm g} (\bm{x}) \cdot {(\bm{p}^{}_{i} - \bm{p}^{\prime}_{j}) } = \mathrm{i} \bm{x}^{}_{\rm vg} \cdot (\bm{p}^{}_{i} - \bm{p}^{\prime}_{j})$, which just represents the E1-type transition with $\bm{x}^{}_{\rm vg} \equiv \left< \rm v \right| \bm{x} \left| \rm g \right>$. The usual electric dipole moment is defined as $\bm{d}^{}_{\rm vg} \equiv - e\, \bm{x}^{}_{\rm vg}$.

The spatial term $\overline{e}  \bm{\gamma} \gamma^{}_{5} e$, recast as $\left< \rm v\right| \bm{\sigma} \left| \rm g \right>$ in the non-relativistic limit, gives rise to a spin-flipped transition that belongs to the M1 type, {where $\bm{\sigma}^{}_{i}$ is the Pauli matrix and $\bm{s}^{}_{i} = \bm{\sigma}^{}_{i} /2$ is just the spin operator}. {For atomic energy levels split by spin inversion, this M1-type matrix element is of order unity. However, in the case of molecular levels, the possible coupling between the electron spin and vibrational or rotational states can introduce suppression factors that depend on the specific type of molecule}. In comparison, the E1-type transition for neutrinos is limited by the Bohr radius $a^{}_{0} = 1/(\alpha m^{}_{e})$, which brings in a suppression factor $m^{2}_{i} a^2_0 \sim 10^{-10}$ in rates compared to the M1 type. 

On the basis of atomic energy eigenstates, the effective Hamiltonian per dipole in the interaction picture, describing the coupling of  fields in the momentum space, turns out to be
\begin{small}
	\begin{eqnarray} \label{eq:HE1} 
		-{H}^{\rm E1}_{\rm I} (t, \bm{x^{}_{d}}) & = &
		\frac{G^{}_{\rm F} C^{}_{ij} }{\sqrt{2}\mathrm{i}} 
		\hat{j}^{ 0}_{i j} (\bm{p}^{}_{i}, \bm{p}^{\prime}_{j}  ) \, \bm{x}^{}_{\rm vg} \cdot (\bm{p}^{}_{i}- \bm{p}^{\prime}_{j})\, \mathrm{e}^{\mathrm{i} (E^{}_{\rm vg} + E^{\prime}_{j} - E^{}_{i} ) t}
		\left|\rm v \right>  \left<\rm g \right| + \bm{d}^{}_{\rm vg} \cdot \bm{\mathcal{E}}(\bm{k}) \, \mathrm{e}^{\mathrm{i} (-E^{}_{\rm vg} + \omega) t } \left|\rm g \right> \left<\rm v \right| + {\rm h.c.} \;,\  \\[0.15cm]  
		\label{eq:HM1} 
		-{H}^{\rm M1}_{\rm I} (t, \bm{x^{}_{d}}) & = &
		\frac{G^{}_{\rm F} C^{}_{ij} }{\sqrt{2}} 
		\hat{\bm{j}}^{}_{i j} (\bm{p}^{}_{i}, \bm{p}^{\prime}_{j}  ) \cdot \bm{\sigma}^{}_{\rm vg}\, \mathrm{e}^{\mathrm{i} (E^{}_{\rm vg} + E^{\prime}_{j} - E^{}_{i}) t  }
		\left|\rm v \right>  \left<\rm g \right|  +  \bm{d}^{}_{\rm vg} \cdot \bm{\mathcal{B}}(\bm{k}) \, \mathrm{e}^{\mathrm{i} (-E^{}_{\rm vg} + \omega) t } \left|\rm g \right> \left<\rm v \right| + {\rm h.c.} \;,\
	\end{eqnarray}
\end{small}
for the E1-type and M1-type interactions, respectively, and we recall that $\omega$  represents the photon energy. 

{The contributing neutrino current operator in the momentum space is given by 
	\begin{align}
		\hat{j}^{\mu}_{i j} = \overline{\nu}^{(+)}_{\bm{p}'_{j}} \gamma^\mu (1-\gamma^{}_{5}) \nu^{(-)}_{\bm{p}^{}_{i}} + \overline{\nu}^{(-)}_{\bm{p}^{}_{i}} \gamma^\mu (1-\gamma^{}_{5}) \nu^{(+)}_{\bm{p}^{\prime}_{j}} \;.
	\end{align}
	Here, the neutrino field operator has been accordingly decomposed into  discrete momentum modes, namely $\nu(t,\bm{x}) = \sum^{}_{\bm{p}} \nu^{(-)}_{\bm{p}} \mathrm{exp}(-\mathrm{i} E t) + \nu^{(+)}_{\bm{p}} \mathrm{exp}(\mathrm{i} E t)$, where the negative and positive frequencies correspond to the absorption and emission of particles, respectively.
	The spatial dependence of field is incorporated in the field operators as $\nu^{(-)}_{\bm{p}} = \hat{a}_{\bm{p}} u(p) \, \mathrm{exp}( \mathrm{i} \bm{p} \cdot \bm{x})$ and $\nu^{(+)}_{\bm{p}} = \hat{b}^\dagger_{\bm{p}} v(p) \, \mathrm{exp}(- \mathrm{i} \bm{p} \cdot \bm{x})$, with $\hat{a}^\dagger$ and $\hat{b}^\dagger$ being the creation operators of neutrinos and antineutrinos, respectively.
	For Majorana neutrinos, we further have a  relation $\hat{a}^\dagger = \hat{b}^\dagger$.
	
	Contracting with external legs of $\nu^{}_{i}$ and $\nu^{}_{j}$, we formally obtain $\left< \left.\left.\bm{p}^{\prime}_{j} \right| \hat{j}^{\mu}_{ij} \right| \bm{p}^{}_{i} \right> = j^{\mu}_{ij}\, \mathrm{exp}[\mathrm{i} (\bm{p}^{}_{i} - \bm{p}^{\prime}_{j}) \cdot \bm{x}]$. The value of the phase-independent current $j^{\mu}_{ij}$ will be different for Dirac and Majorana neutrinos.}
For Dirac neutrinos, the current simply reads 
\begin{align} \label{eq:jijD}
	j^{\mu}_{i j} = \overline{u}^{}_{\bm{p}^{\prime}_{j}} \gamma^\mu (1-\gamma^{}_{5}) {u}^{}_{\bm{p}^{}_{i}} \;.
\end{align}
For Majorana neutrinos, the result is 
\begin{align} \label{eq:jijM}
	j^{\mu}_{i j} = \overline{u}^{}_{\bm{p}^{\prime}_{j}} \gamma^\mu (1-\gamma^{}_{5}) {u}^{}_{\bm{p}^{}_{i}} + \overline{v}^{}_{\bm{p}^{}_{i}} \gamma^\mu (1-\gamma^{}_{5}) {v}^{}_{\bm{p}^{\prime}_{j}} \;,
\end{align}
due to two different ways of contraction. 
For the conversion from $\nu^{}_{3}$ to $\nu^{}_{1}$ in the NO case, the outgoing $\nu^{}_{1}$ will be ultra-relativistic because of $m^{}_{3} \gg m^{}_{1}$. As a result, the interference between  contributions of two different helicities in the Majorana case will be negligible.
Hence, we can work on the Dirac case, keeping in mind that the event number for the Majorana case is simply twice that of the Dirac one.

\subsection{The Rate for the E1 Transition}
The fluorescence rate for the E1 transition can be derived similar to the M1 case in the main text. 
For the E1 transition, the coefficient of $\left| \rm v \right>$ under the perturbation of neutrino fields reads
\begin{align}
	C^{(1)}_{\rm v} (t) =   & \frac{G^{}_{\rm F} C^{}_{ij} }{\sqrt{2} \mathrm{i}} 
	\frac{\hat{j}^{ 0}_{i j}\, \bm{x}^{}_{\rm vg} \cdot (\bm{p}^{}_{i}- \bm{p}^{\prime}_{j})}{E^{}_{\rm vg} + E^{\prime}_{j} - E^{}_{i}} \mathrm{e}^{\mathrm{i}(E^{}_{\rm vg} + E^{\prime}_{j} - E^{}_{i}) t}\;.
\end{align}
With the above equation, the polarization of the medium in response to the neutrino field can be obtained.
The effective Hamiltonian density is found to be
\begin{align} \label{eq:Hnng}
	\mathcal{H}^{\rm E1}_{\nu\nu\gamma} =   & \frac{G^{}_{\rm F} C^{}_{ij} }{-\sqrt{2} \mathrm{i}} 
	\frac{n^{}_{\bm{d}}\, [ \hat{j}^{ 0}_{i j} \,  \bm{x}^{}_{\rm vg} \cdot (\bm{p}^{}_{i}- \bm{p}^{\prime}_{j}) ]\, [\bm{d}^{*}_{\rm vg} \cdot \bm{\mathcal{E}}(\bm{k}^{}_{})] }{E^{}_{\rm vg} + E^{\prime}_{j} - E^{}_{i}} \times \mathrm{e}^{\mathrm{i}(\omega + E^{\prime}_{j} - E^{}_{i} ) t }  +  {\rm h.c.}\;,
\end{align}
which couples two neutrinos and one photon.
For a uniform medium, after integrating over the target volume
we obtain the phase-matching condition.
We also point out a scenario that the target material has a periodic  profile, e.g., $n^{}_{d}(\bm{x}) = n^{0}_{d} [1+ \cos( \bm{D} \cdot \bm{x})]$, similar to the diffraction grating in optics.
In this case, the phase-matching condition will be modified according to $\delta^{(3)}_{}(\bm{p}^{}_{i}-\bm{p}^{\prime}_{j}- \bm{k}^{}_{} \pm \bm{D})$, deviating from the momentum conservation in vacuum.
The scattering will then induce a momentum transfer of magnitude $|\bm{D}|$ to the material.
Noting $\mathrm{i}T = \mathrm{i} \mathcal{M} \cdot (2\pi)^4 \delta^{(4)}_{}({p}^{}_{i}-{p}^{\prime}_{j}- {k})$, the total amplitude is extracted as 
\begin{align}
	\mathcal{M} = &\frac{G^{}_{\rm F} C^{}_{ij} }{\sqrt{2} } 
	\frac{n^{}_{\bm{d}}\, [ {j}^{ 0}_{i j}  \bm{x}^{}_{\rm vg} \cdot (\bm{p}^{}_{i}- \bm{p}^{\prime}_{j}) ]\, [\omega\bm{d}^{*}_{\rm vg} \cdot \bm{\epsilon}^*(\bm{k})] }{-E^{}_{\rm vg} - E^{\prime}_{j} + E^{}_{i}} \;,
\end{align}
where $\bm{\epsilon}$ stands for the polarization vector of the signal photon.
The relation $\bm{\mathcal{E}} = -\partial \bm{A} / \partial t$ has been used in obtaining the above amplitude.

The expression for the rate turns out to be
\begin{align} \label{eq:Gamma}
	\Gamma & \approx  \frac{G^2_{\rm F} |U^*_{e i} U^{}_{e j}|^2 }{2 \pi e^2} \frac{ f^{\rm E1}_{\theta}\, n^2_{\bm{d}} \, |\bm{d}^{}_{\rm vg}|^4 D^{}_{k}\, \omega |\bm{k}|^4  }{(E^{}_{\rm vg} + E^{\prime}_{j} - E^{}_{i})^2 + \gamma^2_{\rm vg}/4}  \;,
\end{align}
where the relation $\bm{x} \equiv -\bm{d} /e$ has been applied, and $f^{\rm E1}_{\theta}$ is the form factor depending on the polarization of dipoles.
The magnitude of $f^{\rm E1}_{\theta}$ can  be adjusted to be of $\mathcal{O}(1)$ by choosing a polarized material or adopting the periodic film structure even if the medium is isotropic.
Assuming a coherence timescale $T^{}_{\rm c} \sim \mathcal{O}(1~{\rm ns})$, the event rate for a target of dimension $V=(3~{\rm m})^3$ is
\begin{align}
	R^{}_{\nu_3 \to \nu_1} &\sim 4\times 10^{-8}~{\rm yr^{-1}} \;, 
\end{align}
if the refractive index is unity, i.e., $|\bm{k}| = \omega$.
Compared to the M1 case, the event number for the E1 transition is too small to be detectable. The primary reason is that the neutrino-induced current is suppressed by a factor related to the Bohr radius, despite the fact that the electromagnetic current in the E1 case is larger than in the M1 case.

{
	\subsection{Slow-light Phenomenon near the Transition Resonance}
	The slow-light phenomenon is essential for mitigating the impact of the momentum dispersion of relic neutrinos. Although there are various mechanisms to achieve a small group velocity for photons, they generally rely on the scattering of photons with the  resonant structure of the medium~\cite{Khurgin2010}. Intuitively, slow light arises from strong resonant absorption by non-relativistic excitations in the medium, yet in a coherent manner, i.e., the quantum phase of the excitation is preserved.
	During this process, a dressed  ``polariton''-like state, which is a collective mixture of photon and excitation, will be formed.
	Therefore, we should expect a strong modification of the refractive index and hence an altered group velocity in our original setup. Here, we aim to analyze the effect and investigate whether it is possible to achieve a small group velocity without the need of additional structures.

	\begin{figure}[t!]
		\begin{center}
			\includegraphics[width=0.4\textwidth]{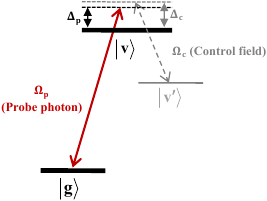}
		\end{center}
		\caption{A two-level system consisting of $\left|\mathrm{g}\right\rangle$ and $\left|\mathrm{v}\right\rangle$, or a three-level system where $\left|\mathrm{v}\right\rangle$ and $\left|\mathrm{v}^\prime\right\rangle$ represent hyperfine-split sublevels. The probe photon couples to the transition dipole  $d_{\mathrm{vg}}$ with a small detuning $\Delta_{\mathrm{p}} = \omega - E_{\mathrm{vg}}$. In the three-level configuration, a control microwave field with Rabi frequency $\Omega_{\mathrm{c}} = d_{\mathrm{v}^\prime \mathrm{v}} \mathcal{E}^{}_{\mathrm{c}}$ is applied to realize electromagnetically induced transparency (EIT). The control field may also be detuned from resonance by $\Delta^{}_{\rm c} = \omega^{}_{\rm c} - E^{}_{\rm v v^\prime}$. In both the two- and three-level systems, the probe photon exhibits a reduced group velocity near the resonance.
		}
		\label{fig:levels}
	\end{figure}
	
	Near the resonance, an unwanted dissipative absorption competes with the slow-light effect, limiting the propagation distance of the slowed light in the medium. This absorption may render the slow-light phenomenon difficult to observe in optical experiments, as the light undergoes significant incoherent attenuation.

	In quantum optics, slow light near resonance is typically observed under the assistance of electromagnetically induced transparency (EIT)~\cite{fleischhauer2005EIT,harris1997electromagnetically,marangos1998electromagnetically,lukin2001controlling,finkelstein2023practical}. This phenomenon requires a three-level system, where an additional energy level is introduced either just above or below the excited state of the original two-level system. When these two upper states are coupled by a strong control field at a frequency close to the level separation (in the visible or microwave range), the medium becomes transparent to a probe photon even at exact resonance.
	At the same time, this photon's dispersion relation will also be significantly modified, leading to a rather reduced group velocity.
	
	In the following, we will investigate two systems depicted in Fig.~\ref{fig:levels}: (i) a simple two-level system formed by $\left| \rm g \right>$ and $\left| \rm v \right>$; (ii) a three-level system including an additional level $\left|\rm v^\prime \right>$ with EIT. The energy separation $E^{}_{\rm v v^\prime}$, which may arise from hyperfine splitting, corresponds to the microwave frequency. For the three-level case, a microwave field will be used to couple $\left| \rm v \right>$ and $\left|\rm v^\prime \right>$~\cite{li2005control,li2009electromagnetically,vogt2018microwave,kosachiov2000efficient,basler2015radio}.
	The preference for microwave is to minimize its possible contamination to our infrared signal, which is far above the microwave frequency. 
	We will quantitatively present their absorption and dispersion profiles, and discuss the implications for our process.
	
	Both absorption and coherent refraction can be described using a complex susceptibility $\chi$. The refractive index is given by  
	$n = \mathrm{Re} (\sqrt{1+\chi})$,
	while the attenuation coefficient due to incoherent absorption is  
	$\alpha^{}_{\rm atten}= 2\omega\,\mathrm{Im} (\sqrt{1+\chi})$ with the corresponding attenuation length $l^{}_{\rm atten} = 1/\alpha^{}_{\rm atten}$.
	For the two-level system, the susceptibility near the resonance is given by~\cite{Khurgin2010}
	\begin{align}
		\chi =  \frac{2  n^{}_{d}\, d^2_{\rm vg}\, E^{}_{\rm vg} }{E^2_{\rm vg} - \omega^2 - \mathrm{i} \omega \gamma^{}_{\rm vg}} \;.
	\end{align}
	For the three-level system, where the virtual states $\left| \rm v \right>$ and $\left|\rm v^\prime \right>$ are coupled with the control field, the susceptibility of a probe photon due to the scattering with the dipole ${d}^{}_{\rm vg}$ will be~\cite{finkelstein2023practical}
	\begin{align}
		\chi =  \frac{\mathrm{i} \,2\,n^{}_{d}\, d^2_{\rm vg}\, \Gamma^{}_{\rm v^\prime g}}{\Gamma^{}_{\rm v^\prime g} \Gamma^{}_{\rm vg} + 4|\Omega^{}_{\rm c}|^2} \;,
	\end{align}
	where $\Gamma^{}_{\rm v^\prime g} \equiv \gamma^{}_{\rm v^\prime g} - \mathrm{i}2  (\Delta^{}_{\rm c} - \Delta^{}_{\rm p})$ and $\Gamma^{}_{\rm v g} \equiv \gamma^{}_{\rm v g} - \mathrm{i} 2  \Delta^{}_{\rm p}$  with $\Delta^{}_{\rm c}\equiv \omega^{}_{\rm c} - E^{}_{\rm v v\prime}$ and $\Delta^{}_{\rm p} \equiv \omega - E^{}_{\rm vg}$ being the one-photon detunings of each field from the level separation. Here, the definitions of decoherence rates $\gamma^{}_{\rm v^\prime g}$ and $\gamma^{}_{\rm v g}$ follow Ref.~\cite{fleischhauer2005EIT}, i.e., this rate will match the spontaneous decay rate in the limit of vanishing purely dephasing rate. The Rabi frequency is $\Omega^{}_{\rm c} = d^{}_{\rm vv^\prime} {\mathcal{E}}^{}_{\rm c}$ with ${\mathcal{E}}^{}_{\rm c}$ being the electric strength of the control field.
	In the limit of vanishing control field, i.e., $\Omega^{}_{\rm c} = 0$, we can recover the expression for the two-level case near the resonance.
	For demonstration, the microwave field strength for the three-level case is taken as $\mathcal{E}^{}_{\rm c} = 2 \times 10^{-2}~{\rm eV}^2$.
	In other units, the chosen value can read as $\mathcal{E}^{}_{\rm c} \approx 3 \times 10^4~{\rm V/m} $ and an  intensity $\mathcal{E}^{2}_{\rm c}/2 \approx 125 ~{\rm W/cm^2}$, which are commonly achievable values in laboratory.
	
	\begin{figure}[t!]
		\begin{center}
			\includegraphics[width=0.4\textwidth]{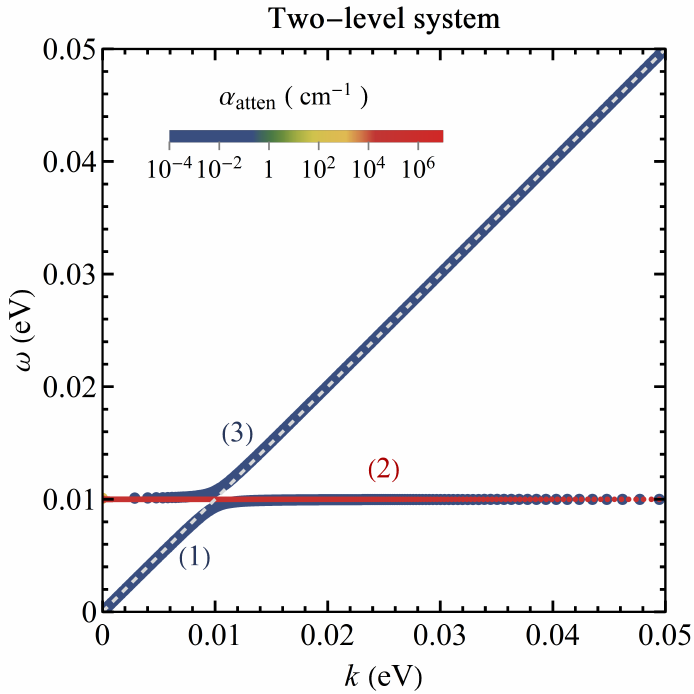} \hspace{0.39cm}
			\includegraphics[width=0.4\textwidth]{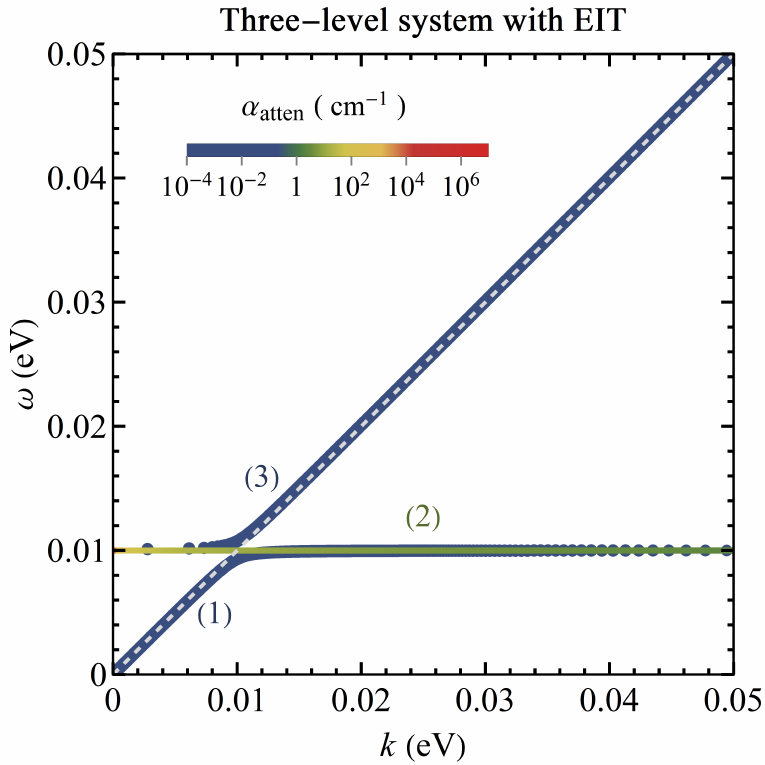}
		\end{center}
		\caption{Dispersion relations of the probe photon in a simple two-level system (left panel) and a three-level system with electromagnetically induced transparency (right panel). The probe photon interacts within an energy interval of $E_{\rm vg} = 0.01~{\rm eV}$. In both cases, the dispersion curves exhibit three connected branches near the resonant frequency, with the color scale indicating the absorption per unit length in the medium. A significant flattening of the dispersion curve, corresponding to a reduced group velocity, is evident near the resonance. The following parameters have been adopted for the demonstration here: $E^{}_{\rm vg} = 10~{\rm meV}$, $d^{}_{\rm vg} = e/(4 m^{}_{e})$, $n^{}_{d} = 6.02 \times 10^{23}~{\mathrm{cm}}^{-3}$ and $\gamma^{}_{\rm vg} = (10~{\rm \mu s})^{-1}$. For the three-level case with EIT, additional parameters are $E^{}_{\rm v  v^\prime} = 10~{\rm \mu eV}$, $d^{}_{\rm v v^\prime } = e/ m^{}_{e}$ and $\gamma^{}_{\rm v^\prime g} = (10~{\rm ms})^{-1}$.
		}
		\label{fig:resonanceSlow}
	\end{figure}

	For the probe photon, the wavevector and group velocity are given by  
	$k = \omega\, n$ and $v^{}_{\rm g} = (\mathrm{d}{k}/\mathrm{d}{\omega} )^{-1}$,
	respectively. Near the resonance at $0.01~{\rm eV}$, the susceptibility varies rapidly with respect to the frequency $\omega$, leading to a highly reduced $v^{}_{\rm g}$. However, without the EIT effect, attenuation also becomes significant in this region.
	Fig.~\ref{fig:resonanceSlow} presents the $\omega$--$k$ dispersion relations without (left panel) and with (right panel) the EIT effect. The attenuation coefficient $\alpha_{\rm atten}$ (whose inverse corresponds to the attenuation length) is indicated by the color along each curve. The dashed line shows the trivial vacuum dispersion.

	\begin{figure}[t!]
		\begin{center}
			\includegraphics[width=0.448\textwidth]{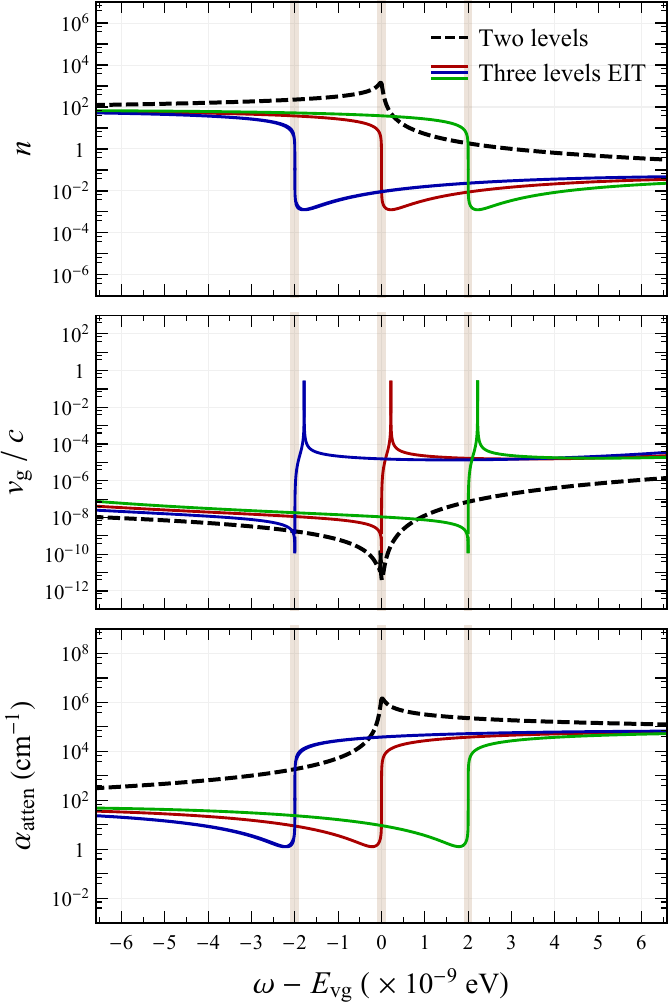} \hspace{0.5cm}
			\includegraphics[width=0.448\textwidth]{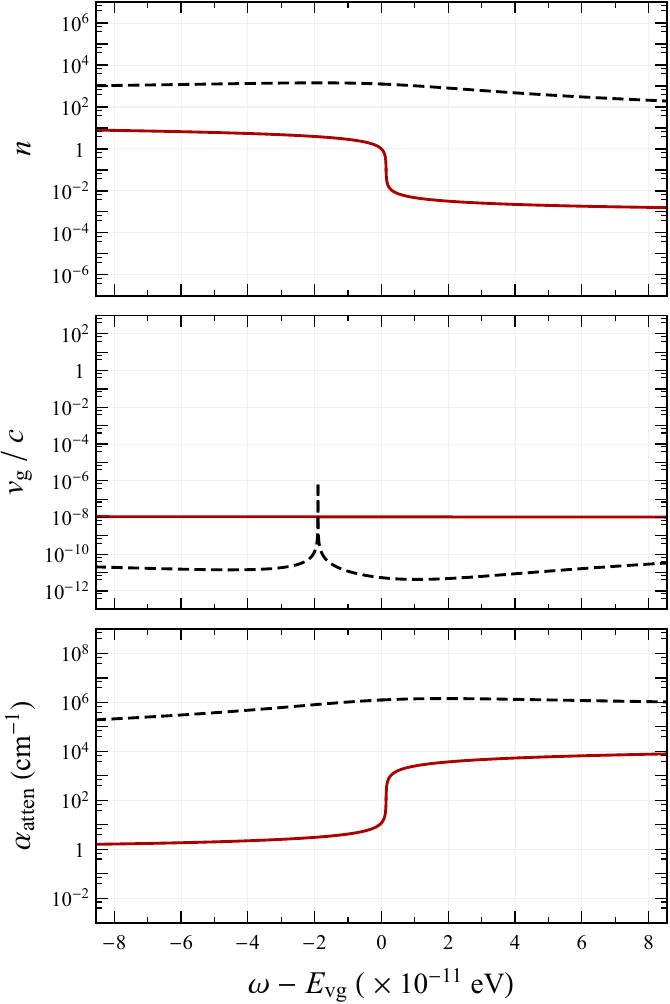}
		\end{center}
		\caption{
			The refractive index $n$ (upper panels), the  group velocity $v_{\rm g}/c$ (middle panels), and the attenuation coefficient $\alpha_{\rm atten}$ (lower panels) as functions of the probe photon detuning $\Delta_{\rm p} = \omega - E^{}_{\rm vg}$. The right panel provides a magnified view of the resonance region in the left panel. In all panels, the dashed curves correspond to the results for the two-level system, while the solid curves represent the three-level system with EIT. Results are shown for three values of the control-field detuning $\Delta_{\rm c}$: $0~{\rm eV}$ (red), $-2\times10^{-9}~{\rm eV}$ (blue), and $2\times10^{-9}~{\rm eV}$ (green).
		}
		\label{fig:resonanceSlowDetails}
	\end{figure}

	We observe that the presence of  resonance splits the dispersion curve into three connected branches within the momentum range of interest, consistent with well-established results in the literature. The first branch originates from the axis origin and rapidly flattens near $\omega \sim E^{}_{\rm vg} = 0.01~{\rm eV}$. The second branch connects to the first at very large $k$ and returns toward $(k, \omega) \sim (0, 0.01)~{\rm eV}$, exhibiting a tiny slope of $\mathrm{d}{\omega}/\mathrm{d}{k} \sim 10^{-7}$. The third branch starts from $(k, \omega) \sim (0, 0.01)~{\rm eV}$ with a small slope and gradually approaches the vacuum dispersion curve as it departs from the resonance. Further comments are given below in order.
	\begin{itemize}[itemsep=2pt,topsep=2pt,leftmargin=5.5mm]
		\item The second branch appears particularly promising for our application due to its minimal resonance detuning $\Delta^{}_{\rm p}$. However, its attenuation length, $l_{\rm atten} \sim 10^{-5}~{\rm cm}$, poses a bottleneck for detecting the initial signal photon and maintaining a large cooperative region for neutrino coherent scatterings. This branch is also recognized as a non-propagating mode characterized by a negative permittivity for the two-level system.
		\item For the first branch, the group velocity at $k = 0.04~{\rm eV}$ remains approximately $v_{\rm g} \sim 4\times 10^{-4}\,c$ with negligible absorption ($l^{}_{\rm atten} > 1~{\rm m}$), but the corresponding frequency is detuned from resonance by $E_{\rm vg} - \omega \sim 10^{-5}~{\rm eV}$, which limits the achievable resonance enhancement even though the slow-light condition is satisfied.
		\item Moving to the right panel for the three-level system, we find that EIT opens a transparency window in the second branch, increasing the attenuation length to $l_{\rm atten} \gtrsim 1~{\rm mm}$ while reducing the group velocity to $v_{\rm g} \lesssim 10^{-8} c$. The detuning at $k^{}_{} = 0.04~{\rm eV}$ is just $10^{-11}~{\rm eV}$.
	\end{itemize}

	Three branches in Fig.~\ref{fig:resonanceSlow} seems overlapping with each other, but in fact rapid changes are happening in a very narrow range near the resonance. To better understand these details, we show in Fig.~\ref{fig:resonanceSlowDetails} the refractive index $n$, the group velocity $v^{}_{\rm g}/c$, and the attenuation strength $\alpha^{}_{\rm atten}$ in the vicinity of the resonance. The left panel covers an energy range of $10^{-8}~{\rm eV}$ around the resonance. The dashed curves stand for the pure two-level case, while the colorful curves represent the three-level case with EIT for different detunings $\Delta^{}_{\rm c}$ of the control field: $0~{\rm eV}$ (red), $-2\times 10^{-9}~{\rm eV}$ (blue) and $2\times 10^{-9}~{\rm eV}$ (green). 
	The detuning clearly shifts the overall spectral features.
	In the right panel, we further zoom into the range of $10^{-10}~{\rm eV}$.
	Within an even smaller detuning window of $|\Delta^{}_{\rm p}| \sim 10^{-11}~{\rm eV}$, the refractive index (proportional to wavevector) drops sharply from $n^{}_{} \sim 5$ ($k \sim 0.05~{\rm eV}$) to $n \sim 1$ ($k \sim 0.01~{\rm eV}$), accompanied by a significant reduction in both group velocity and attenuation coefficient. This rapid behavior just explains the second branch observed in the right panel of Fig.~\ref{fig:resonanceSlow}.

	To conclude this section, the existence of a slow-light mode near the transition resonance is naturally ensured. 
	The resonance enhancement occurs as the curve of neutrino's energy difference intersects with the dispersion curve in the vicinity of the resonance. In the phase space, there always exists a region where the energy conservation and the momentum conservation (the phase-matching condition) hold with a small resonance detuning.
	This is a promising direction worth further investigation, but additional work is needed to apply these effects to specific materials and to understand the neutrino's response within the framework of the Maxwell-Bloch equations describing EIT.

	\subsection{Possible Backgrounds }
	Our detection scheme features signals consisting of one or several photons/phonons at the $\mathcal{O}(10)~{\rm meV}$ scale. This energy threshold is, in principle, achievable using superconducting sensors such as kinetic inductance detectors and superconducting nanowire single-photon detectors,  which are under active development primarily for dark matter searches~\cite{golwala2022novel,Essig:2022dfa,Lyon:2022sza,Cruciani:2022mbb,Gao:2024irf,Temples:2024ntv,Ramanathan:2024hsf,Baudis:2025zyn,TESSERACT:2025tfw,Sandoval:2025mye,Temples:2025xew}.
	In principle, the single-photon or single-phonon detection threshold can approach the Cooper-pair breaking energy, as low as $\sim 1~{\rm meV}$.

	The backgrounds relevant to our signal arise from multiple sources and have been  discussed in the context of light dark matter experiments~\cite{Arvanitaki:2017nhi,Baryakhtar:2018doz,Du:2020ldo,Baudis:2025zyn,Baxter:2022dkm}  as well, where the energy depositions are similarly at the meV level. However, no conclusive answer has yet been provided on the ultimately achievable background rate, given the rapid developments in mitigation strategies such as improved passive shielding and active vetoing.
	Here, we therefore identify the backgrounds most relevant to our scheme and explore potential reduction techniques, including those feasible only in the future.
	Below, we summarize the backgrounds of concern.
	\begin{itemize}[itemsep=2pt,topsep=2pt,leftmargin=5.5mm]
		\item Blackbody radiation. 
		At finite temperatures, thermal photons contribute to the background noise. To suppress this contribution, the material should be maintained at a low temperature in the solid state. The doped ions or molecules embedded in the solid lattice can provide the discrete energy levels that serve as scattering targets. Since the signal photons have energies on the order of tens of meV, a temperature $T \ll 100\,\mathrm{K}$ must be satisfied. The probability to find a photon with energy $\omega$ in the phase space is just given by the Boltzmann distribution $ \mathrm{exp}(-\omega/T)$.
		The rate for those photons to reach the sensors is estimated as~\cite{Arvanitaki:2017nhi,Essig:2019kfe}
		\begin{align}
			\Gamma^{}_{\rm BBR} \sim  \frac{\Delta \omega \, \omega^2}{\pi^2} \mathrm{e}^{-\omega/T} A^{}_{\rm det} \;,
		\end{align}
		where $\Delta \omega$ is the energy resolution, and $A^{}_{\rm det}$ is the area covered by the sensors. Taking $\Delta \omega \sim \omega \sim 10~{\rm meV}$ and $A^{}_{\rm det} \sim 1~{\rm m^2}$, we obtain $\Gamma \approx 1 \times 10^{9}~{\rm s}^{-1}$ for $T = 4~{\rm K}$, and $\Gamma \approx 1.6 \times 10^{-29}~{\rm s}^{-1}$ for $T = 1~{\rm K}$. To control the blackbody background, a temperature of $T \sim 1\,\mathrm{K}$ is thus required, which is readily achievable using standard helium-based cryogenic systems.
		Note that maintaining a low temperature is further motivated by the need to preserve long coherence time of the ensemble.
		\item Cosmic rays and the cosmic microwave background (CMB). 
		Cosmic rays, particularly cosmic muons, induce energetic particle tracks in target materials. While active veto systems can reject muon-induced background tracks, there remains a certain probability that such background signals evade vetoing. Our signal of interest is a single infrared photon or several secondaries, while in most cases, muons induce large energy depositions via cascades, which the veto system can readily reject.
		Since the exact veto efficiency is not yet fully characterized, we instead estimate the minimum efficiency required to suppress this background. Cosmic muon flux decreases exponentially with depth underground. The lowest cosmic-ray muon flux achieved to date is ~$3 \times 10^{-10}~{\rm cm^{-2} \cdot s^{-1}}$ at the China Jinping Underground Laboratory (CJPL)~\cite{JNE:2020bwn}. For a detector with an area of $A^{}_{\rm det} \approx 1~{\rm m}^2$, the corresponding cosmic ray rate is $\Gamma^{}_{\rm CR} = 95~{\rm yr}^{-1}$. A conservative veto efficiency of $99.9\%$ is therefore sufficient to suppress this background.
		This veto can be implemented in multiple ways: by surrounding the target with pure water or scintillators, or by lending the infrared sensor array itself to identifying the highly energetic signatures of track events. If the experiment were conducted in space, the cosmic microwave background (CMB) could contribute additional photon background, but this should not be an issue for underground laboratories. Shielding materials can also efficiently mitigate such photon backgrounds.
		
		\item Radioactivity. 
		There are neutrons and gamma rays originating from natural and cosmogenic radioactive isotopes both within the detector and in the surrounding environment. To  suppress this background, the target may need to be enclosed with a passive shield~\cite{Arvanitaki:2017nhi}, e.g., formed by lead. However, the shielding material itself can introduce additional radioactivity due to impurities, such as $^{238}{\rm U}$.
		In a pessimistic scenario, where the area to shield is about $1~{\rm m}^2$, the resulting radioactivity rate can reach $\Gamma^{}_{\rm RA} \sim 10^{-2}~{\rm s}^{-1}$. In a more optimistic case with a smaller target area of $10~{\rm cm}^2$, the rate is reduced to about $10^{-4}~{\rm s}^{-1}$. This background therefore poses a more significant challenge than those from blackbody radiation or cosmic rays.
		Similar to cosmic rays, radioactive decays typically produce high-energy primary particles that generate numerous secondary photons and phonons, leading to large energy depositions  within a short time window. External veto systems cannot eliminate this background if the radioactivity originates from the target material or its surface sensors themselves. By contrast,  veto schemes based on sensor array itself may effectively reject such events, although their performance depends on the specific target material and detector design. Given the background rates discussed above, the effective observation time for neutrino signals is not expected to be severely reduced, if the relaxation time of background-induced excitations is short.

		\item Afterglow. Long-lived excitations in upper levels of the target medium can be induced similarly by tracks from cosmic rays or radioactivity, later releasing photons known as luminescence~\cite{Derenzo:2016fse,Derenzo:2018plr,Du:2020ldo}. Since our signal photon arises from the near-resonant emission through a higher energy level, we expect background photons of the same energy from the relaxation of the excitation to the ground state.
		If this excitation is metastable, background photon emission can occur randomly long after the track event, which may evade previous veto procedures. This constitutes an irreducible background, even if the sensor's energy resolution is sufficient to reject photons at irrelevant energies.
		To avoid the luminescence background, upon the occurrence of a track activity, a temporal cut to the subsequent time window  could be performed, depending on the lifetime of those excitations. For our case, the typical lifetime of the upper level is estimated to be $\tau \sim 1/(d^2_{\rm vg} E^3_{\rm vg}) \sim 120~{\rm s}$ for M1 transition or $\tau \sim 6~{\rm ms}$ for E1 transition. 
		Furthermore, luminescence originates from the vicinity of the track, with a typical extent of a few mm in our mm-thick disc target. After locating the track, an additional spatial cut on the event position is thus useful to further suppress afterglow contamination.
	\end{itemize}
	
	The discussions above is still preliminary, as the final background rates can be strongly affected, either positively or negatively, by the intrinsic properties of the target material and the performance of superconducting sensors.
	As many light dark matter experiments are advancing along similar lines, significant progress is being made through broad collaborations across multiple fields.
	Nevertheless, unlike the more elusive dark matter, we already understand how neutrinos interact.
	The challenge lies in pushing the experimental frontier far enough to uncover these elusive relics from the Big Bang.
}

\end{document}